\documentclass[preprint,12pt]{elsarticle}
\usepackage[margin=1in]{geometry}
\usepackage{amsmath,amssymb,bm}
\usepackage{booktabs,array,tabularx,multirow}
\usepackage{siunitx}
\usepackage{graphicx}
\usepackage{caption,subcaption}
\usepackage{xcolor}
\usepackage{hyperref}
\usepackage{enumitem}
\usepackage{lineno}
\usepackage[protrusion=true,expansion=false]{microtype} % robust for pdfTeX/MiKTeX bitmap-font setups
\usepackage{longtable}
\usepackage{float}
\usepackage{placeins}
\hypersetup{colorlinks=true, linkcolor=blue, citecolor=blue, urlcolor=blue}
\journal{Composite Structures}
\newcommand{\phif}{\phi_{\mathrm{f}}}
\newcommand{\phiif}{\phi_{\mathrm{if}}}

\newcommand{\kappaf}{\bar{\kappa}_{\mathrm{f}}}
\newcommand{\kappaif}{\bar{\kappa}_{\mathrm{if}}}
\newcommand{\chif}{\chi_{\mathrm{f}}}
\newcommand{\chiif}{\chi_{\mathrm{if}}}

\newcommand{\dd}{\mathrm{d}}
\newcommand{\rev}[1]{\textcolor{black}{#1}}
\newenvironment{revblock}{\begingroup\color{black}}{\endgroup}

\begin{document}
	
	\begin{frontmatter}
		
		\title{A Puck-informed mode-resolved phase-field fatigue framework for
		unidirectional composites}
		
		\author[addr1]{Aamir Dean\corref{cor1}}
		\cortext[cor1]{Corresponding author.}
		\ead{a.dean@sustech.edu}
		\address[addr1]{School of Civil Engineering, College of Engineering, Sudan University of Science and Technology, P.O. Box 72, Khartoum, Sudan}
		
		\begin{abstract}
			Fatigue fracture in unidirectional fibre-reinforced composites is intrinsically
			mode dependent: transverse and off-axis cycling is governed by matrix and
			inter-fibre mechanisms, whereas fibre-aligned cycling activates a distinct
			longitudinal channel with a far larger fracture-energy scale and a different crack
			topology. A model that collapses this behaviour into a single scalar damage
			variable can fit a global stiffness-loss curve but cannot identify which mechanism
			is active. This work develops a Puck-informed, mode-resolved phase-field fatigue
			framework that represents fibre-dominated and matrix/inter-fibre-dominated fatigue
			as two separate channels, each with its own fatigue history, threshold, and
			resistance-degradation law. The central modelling choice is a separation of roles:
			fatigue does not degrade elastic stiffness directly; it lowers the local fracture
			resistance of the active channel, while the corresponding phase field governs the
			actual stiffness loss and crack-path evolution. The formulation is implemented in
			Abaqus/Standard through a compact UMAT--UEL architecture, with one orthotropic
			mechanical routine coupled to two scalar phase-field layers.
			
			The study is verification-first and uses a single fixed IM7/8552 material and
			fatigue card throughout. One-element tests confirm selective channel activation;
			parameter sweeps establish the role of each fatigue constant; and a fixed-amplitude
			study demonstrates the optional channel-wise mean-stress correction. The same fixed
			card is then applied, without any orientation- or geometry-specific tuning, to two
			independent notched configurations -- centred-notch tension (CNT) and open-hole
			tension (OHT) -- at $0^\circ$, $45^\circ$, and $90^\circ$ under both monotonic and
			cyclic loading. In every case the framework reproduces the physically expected
			mode: transverse matrix/inter-fibre cracking at $90^\circ$, off-axis cracking at
			$45^\circ$, and longitudinal matrix splitting with shielded, delayed fibre activation
			at $0^\circ$. The fatigue lives are ordered consistently with these mechanisms: the
			$90^\circ$ and $45^\circ$ configurations fail within \rev{$\sim 10^{3}$ cycles}, whereas
			both $0^\circ$ configurations run out to $2\times10^{5}$ cycles with stable matrix
			splitting and no fibre crack. Supplementary load-controlled and hole-size studies
			show the expected monotonic trends -- lower amplitude and smaller notches delay
			matrix initiation and extend life -- while preserving the governing crack mode.
			Mesh/length-scale and cycle-block convergence checks confirm that the reported
			crack paths and lives are controlled by the regularization length and not by spatial
			or temporal resolution. The results show that a single, fixed, physically
			interpretable mode-resolved card produces consistent, mechanism-correct fatigue
			behaviour across orientation, load level, and notch geometry. The study is
			therefore interpreted as a numerical verification and cross-geometry consistency
			assessment of the proposed formulation, not as a calibrated experimental
			life-prediction claim.
		\end{abstract}
		
		
	\end{frontmatter}
	
	%\linenumbers
	
	\section{Introduction}
	
	The fatigue life of a unidirectional fibre-reinforced composite (UD-FRC) is governed
	by the damage mechanism that is active, not merely by the amount of stiffness that has
	already been lost. A lamina that undergoes rapid transverse matrix cracking under
	transverse or off-axis cyclic loading may remain stable for orders of magnitude more
	cycles when the load is aligned with the fibres, before any fibre-dominated localization
	is reached \citep{talreja1981fatigue,gamstedt1999fatigue}. These mechanisms differ
	in strength scale, fracture energy, characteristic length, and crack topology. A
	computational model that represents cyclic degradation through a single scalar damage
	variable can be tuned to match a measured stiffness-loss curve, yet it remains blind
	to the distinction that governs fatigue life and residual load capacity: it cannot say
	whether the next increment of damage is a matrix split, an inter-fibre crack, or a
	fibre-cutting crack. Predicting orientation-, notch-, and load-dependent fatigue in
	UD composites therefore requires a model that resolves the active mechanism, not only
	its aggregate structural effect. Under transverse and off-axis loading the response is
	dominated by matrix cracking, fibre--matrix debonding, and shear-driven inter-fibre
	fracture; under fibre-aligned loading the lamina may retain most of its longitudinal
	load path while matrix/inter-fibre splitting redistributes the local notch field.
	
	The need for such a distinction is well established in the composite-fatigue
	literature. Early mechanism-based interpretations already showed that fatigue-life
	diagrams and residual-strength evolution for composite laminates cannot be understood
	without identifying the active damage mechanism \citep{talreja1981fatigue,
		reifsnider1982fracture}. Subsequent micromechanical observations in UD carbon-fibre
	reinforced polymers confirmed that nominally similar stiffness-loss trends may arise
	from different damage processes, including distributed fibre breaks, fibre--matrix
	debonding, matrix cracking, and interfacial splitting \citep{gamstedt1999fatigue}.
	Reviews of composite-fatigue modelling further classify existing approaches into
	life-curve, residual-stiffness/residual-strength, and progressive-damage families,
	making clear that a single phenomenological damage variable is usually insufficient
	when mechanism identification is required \citep{degrieck2001review,
		alam2019cfrpreview}. Multiaxial and off-axis fatigue studies also show that ply-level
	normal--shear interactions control fatigue strength and crack initiation, reinforcing
	the need for local mode-sensitive fatigue measures rather than purely global load
	metrics \citep{quaresimin2010multiaxial,carraro2014crackinitiation}. This is
	particularly important for notched laminates and open-hole configurations, where
	interrupted fatigue tests and X-ray CT studies have shown that the sequence of damage
	development, not only the final residual stiffness, controls whether the structure
	runs out or collapses \citep{nixonpearson2013part1,nixonpearson2015openholefatigue}.
	Recent high-cycle fatigue simulations of open-hole laminates likewise emphasize the
	need to track matrix cracks, splits, delamination, local stress-ratio effects, and
	their interaction in a spatially resolved way \citep{hofman2024hcf}.
	
	Phase-field fracture modelling offers a natural basis for this type of spatially
	resolved description because discontinuities are represented by smooth crack fields
	and do not require explicit tracking. Since the variational formulation of brittle
	fracture by Francfort and Marigo \citep{francfort1998}, the numerical regularization
	introduced by Bourdin and co-workers \citep{bourdin2000}, and robust operator-split
	implementations such as that of Miehe et al. \citep{miehe2010}, phase-field approaches
	have been extended to anisotropic fracture, ductile fracture, fatigue crack growth,
	and composite failure. In fatigue, influential formulations have shown how cyclic
	subcritical crack growth can be represented by degrading toughness or fracture
	resistance near the crack process zone, by introducing cycle-dependent crack-growth
	kinetics, or by coupling phase-field fracture to fatigue variables and cyclic
	plasticity \citep{alessi2018,mesgarnejad2019fatiguepf,lo2019fatiguepf,carrara2020,
		seiler2020ductilefatigue,khalil2022generalised,dean2026}. Reviews of composite phase-field
	fracture document the rapid development of diffuse-crack models for laminates,
	including anisotropic crack regularizations, multiple damage mechanisms, and coupling
	to cohesive or interface descriptions \citep{bui2021review}. More broadly, reviews
	of phase-field fatigue show that cyclic degradation is commonly introduced either by
	degrading fracture toughness/resistance or by adding fatigue-dependent crack-driving
	terms \citep{kalina2023overview}. These developments provide the mathematical setting
	for fatigue crack initiation and growth, but they do not by themselves resolve which
	composite failure mechanism is responsible for the observed degradation.
	
	For fibre-reinforced composites, several phase-field formulations have therefore
	introduced additional structure beyond a single isotropic crack field. Early
	laminate-scale phase-field approaches for long-fibre composites demonstrated the
	potential of anisotropic diffuse-crack formulations to capture intralaminar and
	translaminar fracture paths without prescribing crack topology
	\citep{quintanas2019intratrans}. Microstructure-informed phase-field/cohesive models
	have been used to study matrix cracking, fibre--matrix debonding, bridging, and
	R-curve behaviour in fibre-reinforced composites \citep{tan2021microfrc,
		tan2022bridging}. At the ply and laminate scales, Puck-informed multi-phase-field
	formulations exploit the classical distinction between fibre failure and inter-fibre
	failure by introducing separate phase fields, channel-specific fracture energies, and
	channel-specific crack regularizations \citep{dean2020puckmpf,asur2021mpfczm,
		asur2025}. The present work takes this Puck-informed multi-phase-field lineage
	\citep{dean2020puckmpf,asur2025} as its point of departure and extends the same
	mechanistic separation from monotonic fracture to cyclic degradation. Related
	composite phase-field models have also introduced multiple or mode-sensitive crack
	fields to represent competing fibre and matrix failure modes under monotonic loading
	\citep{zhang2022doublepf}. The common lesson from these studies is that crack topology
	is itself mechanistic information: a longitudinal split, an off-axis inter-fibre crack,
	a transverse crack, and a fibre-cutting crack should not be collapsed into one
	undifferentiated scalar fracture variable.
	
	Recent fatigue-oriented phase-field models for fibre-reinforced composites have
	started to incorporate cyclic degradation into this diffuse-crack setting. Li et al.
	proposed a phase-field fatigue framework for fibre-reinforced composites that improves
	matrix-crack paths by accounting for anisotropic fatigue accumulation effects
	\citep{li2024frcfatigue}. Sharma and Singh developed a degradation-informed
	three-dimensional phase-field model for matrix-dominated high-cycle fatigue in
	composite laminates, capturing matrix splitting and shear-driven cracking with little
	fibre rupture in the investigated cases \citep{sharma2026matrixhcf}. In parallel,
	non-phase-field progressive fatigue frameworks based on cohesive-zone models, XFEM,
	continuum damage mechanics, and cycle-jump strategies have demonstrated impressive
	capabilities for matrix cracking, splitting, delamination, local stress-ratio effects,
	and fatigue-life prediction in notched laminates \citep{hofman2024hcf}. These models
	are highly relevant to the present work because they show that composite fatigue must
	be simulated as an evolving crack-process problem. However, they do not formulate
	fatigue accumulation at the level of independent Puck fibre and matrix/inter-fibre
	channels, each with its own threshold, history variable, and fracture-resistance
	degradation law.
	
	A second body of literature is directly relevant to the fibre-aligned response examined
	later in this paper. Experiments on centre-notched UD graphite/epoxy laminates have
	observed longitudinal splitting under tensile loading \citep{wolla1987splitting}, and
	open-hole UD tests have reported fibre-direction splits growing from the hole under
	longitudinal tension \citep{bazhenov1998splitting}. Analytical studies further show
	that such fibre-direction splits can blunt the notch and substantially reduce the local
	stress concentration ahead of the notch tip \citep{liu2016stress}. These observations
	are important because they imply that some forms of matrix/inter-fibre damage are not
	merely precursors to immediate collapse. In fibre-aligned notched plies, longitudinal
	splitting can redistribute the local stress state and delay fibre-dominated fracture.
	A fatigue formulation that cannot distinguish a stable matrix split from a fibre-cutting
	crack cannot express this shielding mechanism in a physically transparent way.
	
	The gap addressed here is therefore specific. Classical fatigue and progressive-damage
	models recognize that composite fatigue is mechanism dependent; Puck-informed
	multi-phase-field models resolve fibre and inter-fibre fracture by mode, but have so
	far been used primarily for monotonic fracture; and recent composite phase-field
	fatigue models introduce cyclic degradation, but not as independent fatigue histories
	assigned to the Puck fibre and matrix/inter-fibre channels. To the author's knowledge,
	no existing formulation combines these ingredients into a single UD-ply fatigue
	framework in which the fibre and matrix/inter-fibre Puck channels possess separate
	fatigue thresholds, accumulation exponents, history variables, and resistance-degradation
	functions, while the actual stiffness loss and crack topology remain governed by
	separate phase fields. In short, the proposed framework is fatigue-capable like
	phase-field fatigue models and mode-resolved like Puck-informed multi-phase-field
	models, but it is both at once: cyclic degradation is assigned to the same physical
	channels that distinguish fibre failure from inter-fibre failure. This combination is
	not merely formal. Without it, a model cannot represent the possibility that a coupon
	is severely matrix/inter-fibre damaged yet remains structurally stable, because it
	cannot separate a stable longitudinal split from a fibre-cutting crack. The present
	work closes this gap by making fatigue accumulation mode-resolved at the same physical
	level at which Puck theory distinguishes fibre failure from inter-fibre failure.
	
	The central hypothesis is that fatigue in UD plies should be represented by two coupled
	but distinct degradation channels. A fibre-dominated channel governs longitudinal and
	fibre-related fracture, whereas a matrix/inter-fibre-dominated channel governs
	transverse and shear-driven matrix/inter-fibre cracking. The fatigue variables
	accumulate channel-wise and reduce the local fracture resistance or threshold of the
	corresponding channel. They do not directly remove elastic stiffness. Actual stiffness
	loss and crack-path evolution remain controlled by the local phase fields $\phif$ and
	$\phiif$. This separation is essential because it allows fatigue to prepare a channel
	for crack growth while preserving a direct physical interpretation of the observed
	degradation state: accumulated cyclic susceptibility is read from the resistance
	variables, whereas actual material degradation is read from the phase fields.
	
	The paper makes five contributions, centred on the mode-resolved fatigue formulation
	and on the split-induced shielding mechanism that this formulation makes observable.
	First, it formulates a mode-resolved fatigue extension of a Puck-informed two-phase-field
	model for UD composites, with channel-specific fatigue histories, thresholds,
	accumulation laws, and fracture-resistance degradation functions. Second, it implements
	the formulation in a compact Abaqus/Standard UMAT--UEL architecture and establishes a
	verification-first programme: one-element checks, parameter-sensitivity curves, and
	structural benchmarks are executed sequentially and interpreted through the phase-field
	variables rather than through raw numerical force or driver spikes. Third, it
	demonstrates, on two independent notched geometries (CNT and OHT) and across
	orientations, that one fixed IM7/8552 card yields consistent, mechanism-correct
	behaviour without per-configuration tuning. Fourth, it shows that the same card reacts
	consistently to load level and to geometric notch severity through load-controlled and
	hole-size studies. Fifth, it isolates and quantifies the physical mechanism behind the
	fibre-aligned runout: longitudinal matrix/inter-fibre splitting shields the fibre
	channel by redistributing the local notch field, so that the fibre phase field remains
	inactive under cyclic loading even though the matrix/inter-fibre phase field has formed
	a stable split. The production results use fixed cycle-block acceleration with explicit
	convergence checks; adaptive cycle-jump control is a natural numerical extension and
	is left to a dedicated algorithmic study.
	
	In one sentence, the claim of this paper is that resolving fatigue into separate,
	physically interpretable fibre and matrix/inter-fibre channels -- each degrading its
	own fracture resistance rather than the elastic stiffness -- is sufficient to reproduce
	the orientation-, load-, and notch-dependent fatigue mechanisms of a UD lamina with one
	fixed parameter card.
	
	The remainder of the paper is organized as follows. Section~\ref{Mode-resolved} the mode-resolved
	phase-field formulation, including the local ply system, channel-wise elastic
	degradation, anisotropic crack-surface regularization, and Puck-informed activation
	rule. Section~\ref{fatigue} introduces the fatigue extension, namely the channel-wise cycle
	measures, fatigue-history accumulation law, and resistance-degradation function.
	Section~\ref{Finite-element} describes the UMAT--UEL implementation, the fixed IM7/8552 material and
	fatigue card, specimen geometries, boundary conditions, and mesh-resolution philosophy.
	Section~\ref{Numerical} defines the serialized numerical programme. Section~\ref{Verification} verifies channel
	separation and parameter controllability in one-element tests. Section~\ref{Centred-notch} and~\ref{Open-hole} present
	the CNT and OHT static and fatigue benchmarks, including load-controlled and hole-size
	sensitivity checks. Section~\ref{Discussion} discusses the mechanism-level interpretation, including
	split-induced shielding, cross-geometry consistency, numerical observables, and
	limitations. Section~\ref{Conclusions} summarizes the conclusions, while the Supplementary Information
	provides additional field-extracted shielding metrics and contour evidence for the
	load-control and geometry-sensitivity studies.
	
	\section{Mode-resolved phase-field formulation}\label{Mode-resolved}
	
	\subsection{Kinematics and local ply system}\label{Kinematics} 
	
	Let $\Omega\subset\mathbb{R}^2$ denote the lamina domain, with displacement field $\bm{u}$ and small-strain tensor
	\begin{equation}
		\bm{\varepsilon}=\mathrm{sym}(\nabla \bm{u}).
	\end{equation}
	All constitutive quantities are evaluated in the local ply coordinate system. The local axis $\bm{e}_1$ is aligned with the fibre direction and $\bm{e}_2$ is transverse to the fibres. For a vertical global loading direction, the input angle $\theta$ in the implementation is related to the paper orientation $\alpha$ by the convention used in the simulations: the nominal $0^\circ$ case corresponds to fibres aligned with the loading direction and is implemented with $\theta=90^\circ$, the $45^\circ$ case with $\theta=45^\circ$, and the $90^\circ$ case with $\theta=0^\circ$.
	
	Two phase-field variables are introduced:
	\begin{equation}
		\phif:\Omega\rightarrow[0,1], \qquad \phiif:\Omega\rightarrow[0,1],
	\end{equation}
	where $\phif$ represents the fibre-dominated crack channel and $\phiif$ represents the matrix/inter-fibre crack channel. In the implementation and post-processing, the most important output variables are SDV41 $=\phif$ and SDV42 $=\phiif$.
	
	\subsection{Channel-wise elastic degradation}
	
	The local orthotropic elastic energy is partitioned into longitudinal, transverse/coupling, and shear-related contributions,
	\begin{equation}
		\psi_0(\bm{\varepsilon}) = \psi_{11}(\bm{\varepsilon})+\psi_{22}(\bm{\varepsilon})+\psi_{12}(\bm{\varepsilon}).
	\end{equation}
	The degraded energy is written as
	\begin{equation}
		\psi^{\mathrm{deg}}(\bm{\varepsilon},\phif,\phiif)=g_{\mathrm{f}}(\phif)\psi_{11}+g_{\mathrm{if}}(\phiif)\psi_{22}+g_{\mathrm{mix}}(\phif,\phiif)\psi_{12},
		\label{eq:degraded-energy}
	\end{equation}
	with
	\begin{equation}
		g_{\mathrm{f}}(\phif)=(1-\phif)^2+\eta, \qquad g_{\mathrm{if}}(\phiif)=(1-\phiif)^2+\eta,
	\end{equation}
	where $\eta$ is a small residual stiffness. The shear transfer is degraded by a smooth minimum-type coupling,
	\begin{equation}
		g_{\mathrm{mix}}=\frac{1}{2}\left[g_{\mathrm{f}}+g_{\mathrm{if}}-\sqrt{(g_{\mathrm{f}}-g_{\mathrm{if}})^2+\epsilon_g^2}\right],
	\end{equation}
	where $\epsilon_g$ is a small smoothing parameter, followed by projection to $[0,1]$ for numerical robustness. This choice reflects that loss of either fibre continuity or matrix/inter-fibre integrity reduces in-plane shear transfer.
	
	The decomposition is taken along the local orthotropic components in the ply frame.
	For the degraded elastic energy and the resulting stiffness, three contributions are
	distinguished: the longitudinal term $\psi_{11}$, the transverse term $\psi_{22}$,
	and the in-plane shear term $\psi_{12}$, degraded respectively by $g_{\mathrm{f}}$,
	$g_{\mathrm{if}}$, and the coupling function $g_{\mathrm{mix}}$ of
	Eq.~\eqref{eq:degraded-energy}. The crack-driving energies that feed the two phase
	fields are formed from the same components but grouped by channel. Using the local
	effective, undegraded stress components, the fibre channel is driven by the longitudinal
	part, $\psi_{\mathrm{f}}=\tfrac{1}{2}\,\sigma_{1}\varepsilon_{1}$, and the
	matrix/inter-fibre channel by the combined transverse and shear part,
	$\psi_{\mathrm{if}}=\tfrac{1}{2}\,(\sigma_{2}\varepsilon_{2}+\sigma_{12}\gamma_{12})$.
	Each driving energy is restricted to its non-negative part, $\langle\cdot\rangle_+$,
	so that compressive normal contributions do not drive tensile cracking; the
	directional (tension versus compression) character of the response is otherwise
	carried by the Puck effort branches of Section~\ref{sec:puck}, not by a spectral strain
	decomposition. This component-wise split keeps each crack channel energetically tied
	to the deformation mode it represents.
	
	In the scalar phase-field solve, each channel is driven by an irreversible history
	field rather than by the instantaneous crack-driving energy, which enforces
	irreversibility of the mechanical driving term,
	\begin{equation}
		H_i^{n+1}=\max\!\big(H_i^{n},\,\zeta_i\,\psi_i^{+}\big),
		\qquad \psi_i^{+}=\langle\psi_i\rangle_+,\qquad i\in\{\mathrm{f},\mathrm{if}\},
		\label{eq:history-field}
	\end{equation}
	where $\zeta_i$ is a channel-wise driving-energy scale. The history field is updated
	only once the corresponding Puck channel has activated (Section~\ref{sec:puck}); before
	activation the channel is held inactive as described there. In all reported simulations
	$\zeta_{\mathrm{f}}=\zeta_{\mathrm{if}}=1$, and the scales are retained only as optional
	numerical parameters.
	
	\subsection{Crack-surface regularization}
	
	The fracture contribution contains two crack-surface densities,
	\begin{equation}
		\Gamma_i(\phi_i)=\int_\Omega \gamma_i(\phi_i,\nabla\phi_i)\,\dd\Omega, \qquad i\in\{\mathrm{f},\mathrm{if}\}.
	\end{equation}
	For an AT2-type regularization,
	\begin{equation}
		\gamma_i(\phi_i,\nabla\phi_i)=\frac{\phi_i^2}{\ell_i}+\ell_i \nabla\phi_i\cdot \bm{A}_i \nabla\phi_i,
		\label{eq:crack-density}
	\end{equation}
	where $\ell_i$ is the characteristic length and $\bm{A}_i$ is a channel-specific
	structural tensor. The associated regularized fracture energy is
	$(G^{\mathrm{eff}}_{\mathrm{c},i}/c_w)\,\Gamma_i$ with $c_w=2$, so that the standard AT2
	crack-surface energy density
	$(G^{\mathrm{eff}}_{\mathrm{c},i}/c_w)(\phi_i^2/\ell_i+\ell_i\nabla\phi_i\cdot\bm{A}_i\nabla\phi_i)$
	is recovered and, for the optimal one-dimensional profile, the surface energy converges
	to the effective Griffith value $G^{\mathrm{eff}}_{\mathrm{c},i}$; in the undamaged
	case $\chi_i=1$, this reduces to the virgin channel toughness $G_{\mathrm{c},i}$.
	The structural tensors $\bm{A}_i$ are anisotropic and set the preferred orientation of
	each regularized crack band. Writing $\bm{n}=(\cos\theta,\sin\theta)$ for the local
	fibre direction and $\bm{m}=(\sin\theta,-\cos\theta)$ for the transverse direction, the
	matrix/inter-fibre tensor is the projector along the fibres,
	$\bm{A}_{\mathrm{if}}=\bm{n}\otimes\bm{n}$, which penalises phase-field gradients in the
	fibre direction and therefore aligns the inter-fibre crack band \emph{with} the fibres,
	reproducing longitudinal splitting. The fibre tensor is the complementary projector,
	$\bm{A}_{\mathrm{f}}=\bm{m}\otimes\bm{m}$, which penalises gradients across the fibres
	and orients the fibre crack band \emph{perpendicular} to them, reproducing
	fibre breaking. Each channel is thus regularized consistently with the crack topology
	it represents, and the orientation follows automatically from the input material angle
	$\theta$. In contour plots the crack is therefore a regularized band, not a zero-thickness line. The band width is governed by $\ell_i$ and should be interpreted as the numerical representation of a fracture/process zone rather than as a physical crack opening.
	
	\subsection{Puck-informed channel activation}\label{sec:puck}
	
	Puck failure theory \citep{puck1998,puck2002} is used to distinguish fibre and
	inter-fibre failure efforts in the local ply system, within the broader tradition of
	mode-separated ply failure criteria such as Hashin's fibre/matrix criterion
	\citep{hashin1980}. Let $F^{\mathrm{raw}}_{\mathrm{f}}$ and
	$F^{\mathrm{raw}}_{\mathrm{if}}$ denote the raw channel efforts obtained from the
	Puck-informed evaluation. The quasi-static model is recovered when fatigue degradation is inactive and the channel resistance functions remain equal to unity. Under fatigue, the raw efforts are modified by channel-wise resistance degradation,
	\begin{equation}
		F^{\mathrm{eff}}_{i}=\frac{F^{\mathrm{raw}}_{i}}{\max(\chi_i,\chi_{\min})}, \qquad i\in\{\mathrm{f},\mathrm{if}\},
		\label{eq:effective-puck}
	\end{equation}
	where $\chi_i$ is the fatigue degradation function and $\chi_{\min}$ is a numerical lower bound.
	Channel activation is handled by an irreversible latch. A channel opens when its
	effective Puck effort first reaches unity; before activation its phase field is held
	at the previously converged value by a penalty acting on the trial nodal field, which
	suppresses both spurious growth and numerical healing of the inactive channel. After
	activation the phase field evolves under its fatigue-degraded resistance, and a
	monotone (irreversibility) penalty prevents any decrease. This makes activation a
	one-way event and keeps the two channels independent at the element level.
	
	\section{Mode-resolved fatigue extension}\label{fatigue}
	
	\subsection{Cycle measures and fatigue-history accumulation}
	
	For constant-amplitude loading with ratio $R$, the converged mechanical state in the
	fatigue step is taken as the cycle maximum, and the corresponding Puck efforts
	$F^{\mathrm{raw}}_{i,\max}$ are evaluated from the effective stress. The cycle minimum
	is obtained by re-evaluating the Puck efforts at the proportional minimum stress
	$\bm{\sigma}_{\min}=R\,\bm{\sigma}_{\max}$, giving $F^{\mathrm{raw}}_{i}(\bm{\sigma}_{\min})$.
	Because the Puck criterion is branch-defined, a proportional stress scaling does not
	always scale the effort monotonically; to keep the proportional-cycle interpretation
	stable, the cycle-minimum effort is floored at $\max(R,0)\,F^{\mathrm{raw}}_{i,\max}$,
	\begin{equation}
		F^{\mathrm{raw}}_{i,\min}=\max\!\big(F^{\mathrm{raw}}_{i}(\bm{\sigma}_{\min}),\;
		\max(R,0)\,F^{\mathrm{raw}}_{i,\max}\big).
	\end{equation}
	This reproduces the intended amplitude for proportional tensile cycles while still
	admitting an explicit compressive contribution for negative $R$. For each channel, the
	fatigue driver is constructed from a representative channel effort amplitude and mean value,
	\begin{equation}
		F_{a,i}=\frac{1}{2}\left|F^{\mathrm{raw}}_{i,\max}-F^{\mathrm{raw}}_{i,\min}\right|,
		\qquad
		F_{m,i}=\frac{1}{2}\left(F^{\mathrm{raw}}_{i,\max}+F^{\mathrm{raw}}_{i,\min}\right),
	\end{equation}
	with $i\in\{\mathrm{f},\mathrm{if}\}$. A mean-load-corrected equivalent measure is then
	\begin{equation}
		\widehat{F}_{i}=\min\!\left(\widehat{F}_{\max},\;
		\frac{F_{a,i}}{\max\!\big(1-\beta_{\mathrm{mean},i}\,\langle F_{m,i}\rangle_+,\;\epsilon\big)}\right),
	\end{equation}
	where $\beta_{\mathrm{mean},i}$ is a channel-wise mean-stress sensitivity parameter,
	$\langle F_{m,i}\rangle_+=\max(F_{m,i},0)$ ensures that only a tensile mean effort
	reduces the denominator, $\epsilon$ is a small lower bound on the denominator, and
	$\widehat{F}_{\max}$ is an upper cap applied for numerical robustness (set well above
	the activation level, so that it never limits the physically relevant range). Both the
	cycle-maximum and cycle-minimum efforts are evaluated from the \emph{effective}
	(undegraded) stress, so that the fatigue driver reflects the applied mechanical cycle
	rather than the already-degraded stress state; degradation enters the fatigue problem
	only through the resistance functions $\chi_i$. The structural CNT and OHT benchmarks in
	this paper are all performed at $R=0.1$ and use the fixed values
	$\beta_{\mathrm{mean},\mathrm{f}}=\beta_{\mathrm{mean},\mathrm{if}}=0$, so that no
	uncalibrated stress-ratio fitting is introduced into the study. The capability of the
	mean-stress term itself is demonstrated separately in the one-element sensitivity
	programme by keeping the cycle amplitude fixed while varying $R$.
	
	The per-cycle fatigue-history increment is
	\begin{equation}
		\Delta \bar{\kappa}_{i,\mathrm{cyc}}=C_{\mathrm{fat},i}\left\langle \widehat{F}_{i}-F_{\mathrm{th},i}\right\rangle_+^{p_{\mathrm{fat},i}},
		\label{eq:fatigue-increment}
	\end{equation}
	with accumulation coefficient $C_{\mathrm{fat},i}$, threshold $F_{\mathrm{th},i}$, exponent $p_{\mathrm{fat},i}$, and Macaulay bracket $\langle x\rangle_+=\max(x,0)$. For an accepted block of $\Delta N$ cycles,
	\begin{equation}
		\bar{\kappa}_{i}^{n+1}=\bar{\kappa}_{i}^{n}+\Delta N\,\Delta \bar{\kappa}_{i,\mathrm{cyc}}.
	\end{equation}
	The accumulated histories are irreversible and channel-specific.
	
	\subsection{Fatigue degradation of fracture resistance}
	
	Fatigue lowers the effective crack resistance through a smooth asymptotic degradation law,
	\begin{equation}
		\chi_i(\bar{\kappa}_i)=
		\begin{cases}
			1, & \bar{\kappa}_i\le \kappa_{T,i},\\[3pt]
			\left(\dfrac{2\kappa_{T,i}}{\bar{\kappa}_i+\kappa_{T,i}}\right)^{a_i}, & \bar{\kappa}_i>\kappa_{T,i},
		\end{cases}
		\label{eq:chi-law}
	\end{equation}
	where $\kappa_{T,i}$ is the transition history and $a_i$ controls the post-threshold degradation rate. The effective fracture energy is then
	\begin{equation}
		G^{\mathrm{eff}}_{\mathrm{c},i}=\max(\chi_i,\chi_{\min})\,G_{\mathrm{c},i}.
	\end{equation}
	
	This separation of roles is the conceptual core of the formulation. Fatigue acts only
	on the fracture resistance: it lowers the effort needed to activate a channel through
	Eq.~\eqref{eq:effective-puck}, and, once the channel is active, it lowers the resistance
	opposing crack growth. Fatigue never removes elastic stiffness on its own. Stiffness is
	lost only through the phase fields $\phif$ and $\phiif$. The consequence is that the
	accumulated fatigue state and the observed material degradation remain separately
	interpretable at every point: one reads how close a channel is to activation from its
	resistance, and how much load capacity has actually been lost from its phase field. A
	single-variable fatigue model conflates these two, and with them the mechanism.
	
	The coupled problem is solved in a staggered manner. The mechanical equilibrium
	subproblem uses the degraded orthotropic energy of Eq.~\eqref{eq:degraded-energy},
	while each scalar phase-field subproblem minimizes the history functional
	\begin{equation}
		\Pi_{\phi}\big(\phi_i;H_i,\bar{\kappa}_i\big)=
		\int_\Omega g_i(\phi_i)\,H_i\,\dd\Omega
		+\int_\Omega \frac{G^{\mathrm{eff}}_{\mathrm{c},i}}{c_w}\,\gamma_i(\phi_i,\nabla\phi_i)\,\dd\Omega,
		\qquad i\in\{\mathrm{f},\mathrm{if}\},
		\label{eq:phasefield-functional}
	\end{equation}
	where $g_i$ denotes the corresponding channel degradation function,
	$g_{\mathrm{f}}$ or $g_{\mathrm{if}}$, $\gamma_i$ is given by
	Eq.~\eqref{eq:crack-density}, $c_w=2$ is the AT2 normalization, and $H_i$ is the
	history field of Eq.~\eqref{eq:history-field}. The fatigue-degraded fracture energy
	$G^{\mathrm{eff}}_{\mathrm{c},i}=\max(\chi_i,\chi_{\min})\,G_{\mathrm{c},i}$ couples the
	accumulated fatigue history $\bar{\kappa}_i$ into the phase-field subproblem through the
	resistance function $\chi_i$ of Eq.~\eqref{eq:chi-law}. The quasi-static two-field
	formulation is recovered when $\chi_{\mathrm{f}}=\chi_{\mathrm{if}}=1$.
	
	\subsection{Irreversibility and numerical interpretation}
	
	The irreversibility conditions are
	\begin{equation}
		\dot{\phif}\ge0, \qquad \dot{\phiif}\ge0, \qquad \dot{\kappaf}\ge0, \qquad \dot{\kappaif}\ge0.
	\end{equation}
	All reported structural damage states are interpreted from $\phif$ and $\phiif$. The
	physical state is read from the phase fields and the smooth structural response, not
	from instantaneous nodal quantities. Near localization, a displacement-controlled solve can
	produce sharp reaction-force or channel-effort transients as the tangent stiffness changes
	abruptly within a single increment. Such transients are admitted as physical only when they
	are corroborated by smooth load--displacement or compliance evolution and by a consistent
	phase-field pattern; otherwise they are treated as numerical and excluded from reported
	strengths and lives. This rule is applied uniformly to every case, so that no result depends
	on a single-increment force value.
	
	\section{Finite-element implementation and material card}\label{Finite-element}
	
	\subsection{UMAT--UEL architecture}
	
	The formulation is implemented in Abaqus/Standard using a compact staggered UMAT--UEL
	architecture. The mechanical displacement problem is handled by a UMAT. Within each
	increment it evaluates the degraded orthotropic stiffness, computes the stresses, updates
	the channel-wise fatigue histories, and evaluates the Puck-informed channel efforts. The
	resulting state variables are then exported to the two phase-field layers. The two scalar
	phase-field equations are solved by two overlay UEL layers. The fibre phase-field layer and
	the inter-fibre phase-field layer use distinct scalar degrees of freedom, preventing the two
	crack systems from sharing the same Abaqus field variable. The implementation allocates 80
	state variables and uses 59 user-material constants. The important output variables for the
	present paper are SDV41 ($\phif$) and SDV42 ($\phiif$).
	
	The analysis is split into a preload step and a fatigue step. In static simulations the fatigue flag is inactive. In fatigue simulations the preload is first applied without fatigue accumulation to verify that the selected maximum load or displacement is subcritical. The subsequent fatigue step keeps the maximum cyclic load or displacement fixed while the cycle counter advances through cycle blocks. Unless otherwise stated, the fatigue ratio is $R=0.1$.
	
	Within each load or cycle block the mechanical and phase-field subproblems are advanced
	in a staggered manner, each using the converged state variables from the previous
	substep. Robustness is monitored through the Abaqus residual convergence of every
	accepted increment, the absence of phase-field healing (enforced by the irreversibility
	penalty), monotonic growth of the fatigue histories, and consistency of the phase-field
	pattern across accepted increments. Cases that show isolated reaction-force or
	channel-effort spikes without a corresponding smooth structural response or stable
	phase-field evolution are treated as numerical and are not used to define reported
	strengths or lives, as described in Section~\ref{sec:spikes}.
	
	\subsection{IM7/8552 material and fatigue parameters}
	
	All structural simulations use the same fixed IM7/8552 material card, summarized in Tables~\ref{tab:elastic-strength} and \ref{tab:fatigue-card}. No orientation-specific recalibration is used.
	
	\begin{table}[!htbp]
		\centering
		\caption{Fixed IM7/8552 elastic, strength, and phase-field parameters used in all simulations.}
		\label{tab:elastic-strength}
		\begin{tabular}{ll S[table-format=6.4] l}
			\toprule
			Group & Parameter & {Value} & Unit \\
			\midrule
			Elasticity & $E_{11}$ & 171420 & MPa \\
			& $E_{22}$ & 9080 & MPa \\
			& $G_{12}$ & 5290 & MPa \\
			& $\nu_{12}$ & 0.32 & -- \\
			& $\nu_{21}$ & 0.0169502 & -- \\
			\midrule
			Strength & $R_{1T}$ & 2326 & MPa \\
			& $R_{2T}$ & 62.3 & MPa \\
			& $R_{1C}$ & 1200 & MPa \\
			& $R_{2C}$ & 200 & MPa \\
			& $R_{12}$ & 92.3 & MPa \\
			\midrule
			Phase field & $\ell_{\mathrm{f}}$ & 1.5 & mm \\
			& $\ell_{\mathrm{if}}$ & 1.0 & mm \\
			& $G_{\mathrm{c,f}}$ & 106.3 & N/mm \\
			& $G_{\mathrm{c,if}}$ & 0.277 & N/mm \\
			\midrule
			Puck constants & $p_{21}^{+}$ & 0.35 & -- \\
			& $p_{21}^{-}$ & 0.30 & -- \\
			& $p_{22}^{-}$ & 0.25 & -- \\
			\bottomrule
		\end{tabular}
	\end{table}
	
	\begin{table}[!htbp]
		\centering
		\caption{Common structural fatigue card. The same values are used for CNT and OHT simulations. Mean-stress coefficients are fixed at zero for the present $R=0.1$ study. All fatigue parameters are dimensionless in the present nondimensional effort formulation.}
		\label{tab:fatigue-card}
		\begin{tabular}{ll S[table-format=1.2e-1]}
			\toprule
			Channel & Parameter & {Value} \\
			\midrule
			Fibre & $C_{\mathrm{fat,f}}$ & 1.0e-3 \\
			& $p_{\mathrm{fat,f}}$ & 2.5 \\
			& $F_{\mathrm{th,f}}$ & 0.12 \\
			& $\kappa_{T,\mathrm{f}}$ & 1.0 \\
			& $a_{\mathrm{f}}$ & 0.45 \\
			\midrule
			Inter-fibre & $C_{\mathrm{fat,if}}$ & 8.0e-3 \\
			& $p_{\mathrm{fat,if}}$ & 1.5 \\
			& $F_{\mathrm{th,if}}$ & 0.04 \\
			& $\kappa_{T,\mathrm{if}}$ & 1.0 \\
			& $a_{\mathrm{if}}$ & 0.45 \\
			\midrule
			Mean stress & $\beta_{\mathrm{mean},\mathrm{f}}$ & 0.0 \\
			& $\beta_{\mathrm{mean},\mathrm{if}}$ & 0.0 \\
			\bottomrule
		\end{tabular}
	\end{table}
	
	The fatigue parameters in Table~\ref{tab:fatigue-card} should be understood as a fixed
	numerical demonstration card rather than an experimentally identified IM7/8552 fatigue
	calibration. They were selected after the one-element verification and sensitivity
	studies to produce stable channel separation, matrix/inter-fibre fatigue evolution on
	an accessible cycle scale, and a subcritical fibre channel under the selected
	fibre-aligned fatigue amplitudes. Once selected, the same card is held fixed for all
	CNT, OHT, load-level, and hole-size studies; no parameter is re-tuned by orientation,
	geometry, or load case. Experimental identification of a quantitative IM7/8552 fatigue
	card is a separate task and is outside the scope of the present demonstration.
	
	\subsection{Specimen geometries, boundary conditions, and loading}
	\label{sec:geometry}
	
	All specimens are flat rectangular coupons of width $W=20$~mm, height $H=40$~mm
	(gauge aspect ratio $H/W=2$), and unit nominal thickness $t=1$~mm, modelled as a
	single homogenized ply under plane-stress conditions. The origin is taken at the
	specimen centre, so that $x\in[-10,10]$~mm and $y\in[-20,20]$~mm. Two notch families
	are considered, summarized in Table~\ref{tab:geometry}. The centred-notch tension
	(CNT) specimen contains a central horizontal slit of total length $2a=4$~mm
	($2a/W=0.20$), introduced as a sharp seam through the mid-height of the coupon. The
	open-hole tension (OHT) specimen contains a central circular hole of diameter $D=4$~mm
	($D/W=0.20$); the hole diameter is varied to $D=2.5$ and $5.5$~mm only in the
	geometric sensitivity study of Section~\ref{sec:oht-hole-size-sensitivity}. Nominal stresses are reported on the gross
	section $Wt$, and net-section stresses on the remaining ligament, $(W-2a)\,t$ for CNT
	and $(W-D)\,t$ for OHT.
	
	The fibre orientation is set entirely through the local material angle $\theta$ and not
	through the mesh, so the same finite-element mesh is reused for all orientations of a
	given geometry. Following the convention of Section~\ref{Kinematics}, the nominal $0^\circ$
	(fibre-aligned) coupon uses $\theta=90^\circ$, the $45^\circ$ coupon uses
	$\theta=45^\circ$, and the $90^\circ$ (transverse) coupon uses $\theta=0^\circ$.
	
	Loading is applied as tension along the $y$ (height) direction. The bottom edge
	($y=-20$~mm) is constrained in the loading direction ($u_y=0$), and a single node on
	that edge is additionally pinned in $x$ to remove the rigid-body translation without
	over-constraining the transverse contraction. The top edge ($y=+20$~mm) carries the
	applied load: a prescribed vertical displacement $u_y=U_{\max}$ in the
	displacement-controlled analyses, or a prescribed resultant force $F_{\max}$ in the
	load-controlled analyses of Sections~8.3--8.4. Each analysis is split into a preload
	step, in which the maximum load or displacement is reached without fatigue
	accumulation and the state is verified to be subcritical, followed by a fatigue step
	in which the maximum cyclic level is held fixed at ratio $R=0.1$ while the cycle
	counter advances through fixed accepted blocks of $\Delta N$ cycles.
	
	Each element is represented three times on a shared connectivity: a four-node
	plane-stress continuum element (CPS4) carries the mechanical degrees of freedom, and
	two overlaid four-node scalar user elements carry the fibre and inter-fibre phase
	fields on distinct degrees of freedom (DOF~11 and DOF~12, respectively). The mechanical
	and phase-field layers therefore share nodes and integration points but never share a
	field variable.
	
	\begin{table}[!htbp]
		\centering
		\small
		\caption{Specimen geometries, boundary conditions, and discretization. Both
		families share the same coupon envelope; only the central feature and the mesh
		size differ. The hole diameter is varied only in the D7 sensitivity study
		(Section~\ref{sec:oht-hole-size-sensitivity}).}
		\label{tab:geometry}
		\begin{tabularx}{\textwidth}{p{0.30\textwidth}XX}
			\toprule
			Quantity & CNT & OHT \\
			\midrule
			Width $W$ (mm) & 20 & 20 \\
			Height $H$ (mm) & 40 & 40 \\
			Thickness $t$ (mm) & 1 & 1 \\
			Central feature & horizontal slit, $2a=4$~mm & circular hole, $D=4$~mm \\
			Feature ratio & $2a/W=0.20$ & $D/W=0.20$ \\
			Net ligament & $W-2a=16$~mm & $W-D=16$~mm \\
			Mesh & uniform Cartesian & uniform Cartesian, projected hole \\
			Element size $h$ (mm) & 0.25 & 0.20 \\
			$\ell_{\mathrm{if}}/h$, $\ell_{\mathrm{f}}/h$ & 4, 6 & 5, 7.5 \\
			Element type & \multicolumn{2}{c}{CPS4 mechanical $+$ two scalar UEL phase-field overlays} \\
			Bottom edge ($y=-20$) & \multicolumn{2}{c}{$u_y=0$; one node additionally pinned in $x$} \\
			Top edge ($y=+20$) & \multicolumn{2}{c}{prescribed $u_y=U_{\max}$ (disp.\ control) or $F_{\max}$ (load control)} \\
			Cyclic ratio $R$ & \multicolumn{2}{c}{0.1} \\
			\bottomrule
		\end{tabularx}
	\end{table}
	
	\subsection{Mesh and convergence philosophy}
	
	The crack band must be resolved by several elements across the relevant length scale. The CNT simulations use a uniform square mesh with $h=0.25$ mm, giving $\ell_{\mathrm{if}}/h=4$ and $\ell_{\mathrm{f}}/h=6$. The final OHT simulations use a uniform Cartesian $h=0.20$ mm mesh with a projected circular hole, giving $\ell_{\mathrm{if}}/h=5$ and $\ell_{\mathrm{f}}/h=7.5$. The OHT mesh was selected after rejecting earlier pilot meshes with visible topology concerns around the hole. The final OHT mesh removes the O-grid refinement patch and uses the same topology for all orientations.
	
	For the OHT static cases, a direct comparison between the pilot $h=0.25$ mm mesh and the final $h=0.20$ mm mesh showed moderate changes in load level only, summarized in Table~\ref{tab:convergence}. More importantly, the governing crack mechanisms remained unchanged. The slightly larger $0^\circ$ sensitivity is consistent with the more localized, overload-driven nature of the fibre-aligned response and does not affect any quantitative claim, since the $0^\circ$ static value is reported only as a representative load level and is not used for cross-geometry strength comparison (Section~\ref{sec:spikes}). For the matrix-dominated $45^\circ$ and $90^\circ$ cases, which carry the quantitative comparisons, the mesh sensitivity is below $4\%$ and the governing mode is unchanged. A separate cycle-block convergence check with $\Delta N=10$, 25, and 50 cycles was conducted for representative CNT and OHT fatigue cases. Fatigue lives varied by approximately 2--4\% and the crack paths and governing channels were unchanged. The final simulations therefore use fixed cycle-block acceleration with documented convergence rather than introducing adaptive cycle jumping as a primary manuscript claim.
	
	\begin{table}[!htbp]
		\centering
		\small
		\caption{Mesh sensitivity of the OHT static load level between the pilot $h=0.25$~mm mesh and the final $h=0.20$~mm mesh, reported as the relative change in the extracted load level. The governing crack mode is unchanged in every case. The larger $0^\circ$ change reflects the localized, overload-driven fibre-aligned response and does not affect any quantitative claim, since the $0^\circ$ value is used only as a representative load level.}
		\label{tab:convergence}
		\begin{tabular}{l S[table-format=2.1] l}
			\toprule
			Case & {Change, $h$: 0.25$\to$0.20 mm (\%)} & Governing mode \\
			\midrule
			OHT00 & 7.7 & longitudinal matrix splitting \\
			OHT45 & 3.8 & off-axis matrix/inter-fibre \\
			OHT90 & 2.5 & transverse matrix/inter-fibre \\
			\bottomrule
		\end{tabular}
	\end{table}
	
	\section{Numerical programme}\label{Numerical}

	The simulation campaign is organized into sequential verification, benchmark, and supplementary sensitivity studies. The benchmarks are numerical consistency tests performed with one fixed parameter card: no experimental fitting, orientation-specific calibration, or geometry-specific recalibration is performed in this work.
	\begin{enumerate}[label=D\arabic*., leftmargin=1.5cm]
		\item One-element verification of fibre and matrix/inter-fibre channel separation.
		\item One-element fatigue sensitivity sweeps for accumulation coefficient, threshold, exponent, transition scale, and post-threshold degradation parameter.
		\item[D2b.] One-element fixed-amplitude mean-stress sensitivity check for the optional channel-wise mean-stress correction.
		\item CNT static benchmarks at $0^\circ$, $45^\circ$, and $90^\circ$.
		\item CNT fatigue benchmarks at $0^\circ$, $45^\circ$, and $90^\circ$.
		\item OHT static benchmarks at $0^\circ$, $45^\circ$, and $90^\circ$.
		\item OHT fatigue benchmarks at $0^\circ$, $45^\circ$, and $90^\circ$.
		\item[D6b.] OHT90 load-controlled amplitude check at 70\%, 60\%, and 50\% of static OHT90 capacity.
		\item[D7.] OHT90 geometric notch-size sensitivity with hole diameters $D=2.5$, 4.0, and 5.5 mm under static and load-controlled fatigue conditions.
	\end{enumerate}
	This ordering prevents later structural results from being interpreted before channel separation, parameter controllability, static crack paths, and fatigue evolution have been checked.
	
	\section{Verification and parameter sensitivity}\label{Verification}
	
	\subsection{One-element channel separation}
	
	The one-element tests verify that the two fatigue channels can activate independently. A fibre-dominated case activates only the fibre phase field: $\phif$ reaches unity, $\chif$ decreases to approximately 0.505, and $\phiif$ remains zero. A matrix/inter-fibre dominated case activates only the inter-fibre phase field: $\phiif$ reaches unity, $\chiif$ decreases to approximately 0.255, and $\phif$ remains zero. These results establish that the formulation does not contain an unintended single global fatigue variable hidden inside the implementation.
	
	\begin{figure}[tbp]
		\centering
		\includegraphics[width=\linewidth]{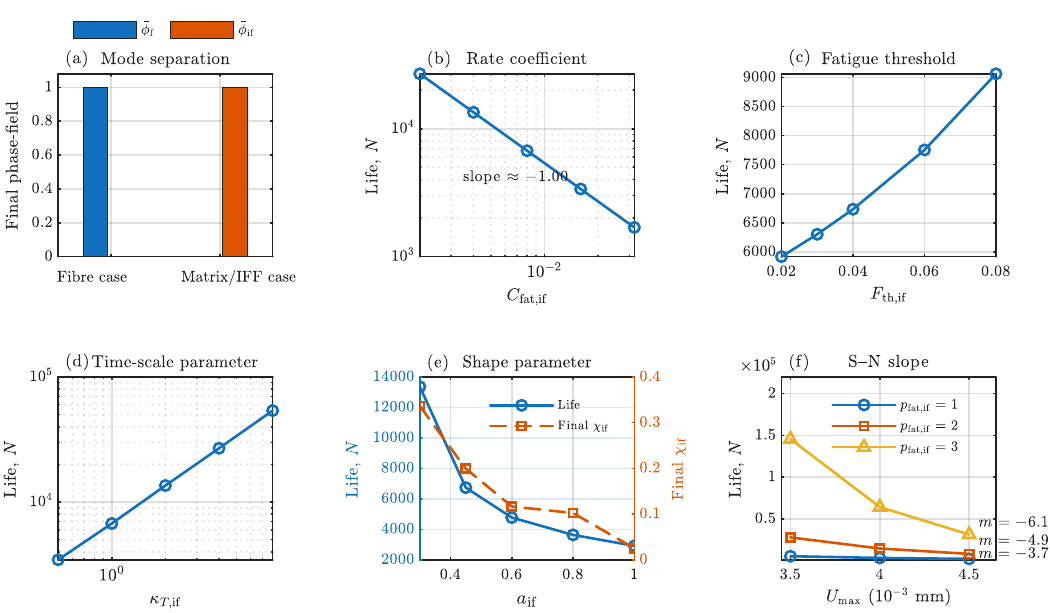}
		\caption{One-element verification and parameter sensitivity of the mode-resolved fatigue formulation. Panel (a) verifies channel separation: the fibre-dominated one-element case activates only the fibre phase-field, whereas the matrix/inter-fibre-dominated case activates only the matrix/inter-fibre phase-field. Panel (b) shows the near-inverse dependence of fatigue life on the matrix/inter-fibre rate coefficient \(C_{\mathrm{fat},if}\). Panel (c) shows that increasing the threshold \(F_{\mathrm{th},if}\) delays matrix/inter-fibre fatigue activation. Panel (d) shows the approximately proportional increase in life with the time-scale parameter \(\kappa_{T,if}\). Panel (e) illustrates the influence of the degradation-shape parameter \(a_{\mathrm{if}}\) on both life and final resistance loss. Panel (f) shows that increasing \(p_{\mathrm{fat},if}\) steepens the displacement--life response, demonstrating direct control of the fatigue slope.}
		\label{fig:d1-d2-verification}
	\end{figure}
	
	\subsection{Parameter sensitivity}
	
	The sensitivity study confirms that each fatigue parameter has the intended role. Increasing $C_{\mathrm{fat,if}}$ accelerates accumulation almost inversely: over the tested range, life scales approximately as $N\propto C_{\mathrm{fat,if}}^{-1}$ with a fitted slope of about $-0.997$ in log--log coordinates. Increasing the threshold $F_{\mathrm{th,if}}$ delays fatigue activation and increases life. Increasing $\kappa_{T,\mathrm{if}}$ shifts the onset of resistance degradation and gives an approximately proportional increase in life. Increasing $a_{\mathrm{if}}$ makes the post-threshold resistance degradation steeper and shortens life. Finally, increasing the exponent $p_{\mathrm{fat,if}}$ steepens the S--N slope: representative log--life slopes become approximately $-3.7$, $-4.9$, and $-6.1$ for $p=1$, 2, and 3, respectively.
	
	\begin{table}[!htbp]
		\centering
		\small
		\caption{Representative D2 sensitivity trends for the matrix/inter-fibre one-element fatigue test. Values are rounded because the table is intended to document parameter roles rather than serve as calibration data.}
		\label{tab:d2-sensitivity}
		\begin{tabularx}{\textwidth}{p{0.18\textwidth}p{0.18\textwidth}Xp{0.22\textwidth}}
			\toprule
			Study & Variation & Observed trend & Interpretation \\
			\midrule
			$C_{\mathrm{fat,if}}$ & $2\times10^{-3}$--$3.2\times10^{-2}$ & life decreases from $2.69\times10^{4}$ to $1.69\times10^{3}$ & accumulation rate \\
			$F_{\mathrm{th,if}}$ & 0.02--0.08 & life increases from $5.90\times10^{3}$ to $9.10\times10^{3}$ & fatigue threshold \\
			$\kappa_{T,\mathrm{if}}$ & 0.5--8.0 & life increases from $3.45\times10^{3}$ to $5.38\times10^{4}$ & onset of resistance loss \\
			$a_{\mathrm{if}}$ & 0.30--1.00 & life decreases from $1.34\times10^{4}$ to $2.91\times10^{3}$ & post-threshold degradation shape \\
			$p_{\mathrm{fat,if}}$ & 1, 2, 3 & S--N slope steepens & amplitude sensitivity \\
			\bottomrule
		\end{tabularx}
	\end{table}
	
	\subsection{Mean-stress capability at fixed cycle amplitude}
	
	The structural fatigue campaign is intentionally restricted to $R=0.1$ with $\beta_{\mathrm{mean},\mathrm{f}}=\beta_{\mathrm{mean},\mathrm{if}}=0$. Nevertheless, the formulation contains an optional channel-wise mean-stress correction, and its behaviour should be transparent. A compact D2b one-element sensitivity study was therefore added to demonstrate the capability without changing the fixed CNT/OHT card. The study uses a matrix/inter-fibre dominated one-element test, keeps the displacement amplitude fixed at
	\begin{equation}
		U_a=\frac{U_{\max}-U_{\min}}{2}=0.0016\,\mathrm{mm},
	\end{equation}
	and varies the load ratio $R=U_{\min}/U_{\max}$ through $0.1$, $0.3$, and $0.5$. The corresponding maximum and minimum displacements are
	\begin{equation}
		U_{\max}=\frac{2U_a}{1-R}, \qquad U_{\min}=R U_{\max}.
	\end{equation}
	Thus the amplitude is held constant while the tensile mean level increases.
	
	Two sets were evaluated. In the control set, $\beta_{\mathrm{mean},\mathrm{if}}=0$, so the equivalent fatigue driver is amplitude controlled. The final equivalent inter-fibre driver remains essentially constant, $\widehat{F}_{\mathrm{if}}\approx0.2345$, for all three $R$ values. The resistance-degradation history is also almost unchanged: the cycle at which $\chi_{\mathrm{if}}<0.5$ remains close to $1.22\times10^{4}$ cycles. The earlier phase-field onset observed at larger $R$ in the control set is therefore not caused by mean-stress amplification, but by the larger maximum cyclic state entering the Puck-based activation gate.
	
\begin{figure}[tbp]
		\centering
		\includegraphics[width=\linewidth]{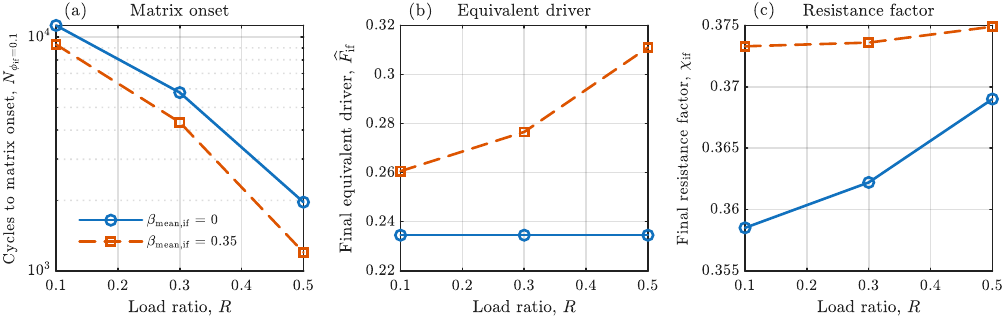}
		\caption{Mean-stress sensitivity of the matrix/inter-fibre fatigue channel at constant displacement amplitude. Panel (a) shows the cycle at which the matrix phase-field reaches \(\phi_{\mathrm{if}}>0.10\), panel (b) shows the final equivalent inter-fibre fatigue driver, and panel (c) shows the final inter-fibre resistance factor. With \(\beta_{\mathrm{mean},if}=0\), the equivalent fatigue driver remains nearly independent of the load ratio \(R\). With \(\beta_{\mathrm{mean},if}=0.35\), increasing tensile mean stress increases the equivalent driver and accelerates resistance degradation and phase-field activation.}
		\label{fig:d2b-meanstress}
	\end{figure}

	In the active set, $\beta_{\mathrm{mean},\mathrm{if}}=0.35$. The equivalent driver then increases with tensile mean level, from 0.2606 at $R=0.1$ to 0.3110 at $R=0.5$. This accelerates the inter-fibre fatigue accumulation and shifts both the resistance degradation and phase-field onset to earlier cycles. Table~\ref{tab:d2b-meanstress} and Figure~\ref{fig:d2b-meanstress} summarize the result. The purpose of this study is not a calibrated stress-ratio validation, but a controlled demonstration that the optional mean-stress term has the expected qualitative effect.
	
	\begin{table}[!htbp]
		\centering
		\small
		\caption{D2b one-element mean-stress sensitivity at fixed displacement amplitude $U_a=\SI{0.0016}{\milli\meter}$. The control set keeps $\beta_{\mathrm{mean},\mathrm{if}}=0$; the active set uses $\beta_{\mathrm{mean},\mathrm{if}}=0.35$. The structural CNT/OHT benchmarks retain $\beta_{\mathrm{mean},\mathrm{f}}=\beta_{\mathrm{mean},\mathrm{if}}=0$. Cycle counts are integer numbers of cycles.}
		\label{tab:d2b-meanstress}
		\begin{tabular}{l S[table-format=1.1] S[table-format=1.4] S[table-format=5.0] S[table-format=5.0] S[table-format=5.0]}
			\toprule
			Set & {$R$} & {$\widehat{F}_{\mathrm{if}}$} & {$N(\phi_{\mathrm{if}}>0.10)$} & {$N(\phi_{\mathrm{if}}>0.98)$} & {$N(\chi_{\mathrm{if}}<0.50)$} \\
			\midrule
			$\beta_{\mathrm{mean},\mathrm{if}}=0$ & 0.1 & 0.2345 & 11219 & 11225 & 12330 \\
			& 0.3 & 0.2345 & 5789 & 5803 & 12216 \\
			& 0.5 & 0.2345 & 1966 & 1972 & 12224 \\
			\midrule
			$\beta_{\mathrm{mean},\mathrm{if}}=0.35$ & 0.1 & 0.2606 & 9293 & 9298 & 10221 \\
			& 0.3 & 0.2766 & 4318 & 4342 & 9120 \\
			& 0.5 & 0.3110 & 1199 & 1205 & 7439 \\
			\bottomrule
		\end{tabular}
	\end{table}
	
	\section{Centred-notch tension benchmarks}\label{Centred-notch}
	
	\subsection{Static CNT response}
	
	The CNT specimen provides a clean notched benchmark with a uniform mesh and three orientations. The static results are summarized in Table~\ref{tab:cnt-static}. CNT00 exhibits longitudinal matrix/inter-fibre splitting first, followed by delayed localized fibre-dominated phase-field activation under overload. CNT45 is governed by an off-axis inter-fibre crack. CNT90 is governed by a transverse inter-fibre crack with no meaningful fibre damage. The CNT00 static response is reported using the full-displacement overload run and a representative smoothed peak before the post-localization spike regime to avoid interpreting post-localization numerical spikes as material strength.
	
	\begin{table}[!htbp]
		\centering
		\small
		\caption{CNT static results. The $45^\circ$ and $90^\circ$ values are governed by the same matrix/inter-fibre mode in both geometries and are reported as strengths; for the $0^\circ$ case the reported value is a representative load level taken from the full-displacement overload continuation (used to expose delayed fibre-channel activation), not an ultimate-strength definition directly comparable to the other cases or to the open-hole geometry (see Section~\ref{sec:spikes}).}
		\label{tab:cnt-static}
		\begin{tabularx}{\textwidth}{l S[table-format=5.0] S[table-format=1.3] S[table-format=4.1] S[table-format=4.1] X}
			\toprule
			Case & {Reported force} & {$U$ at peak} & {Nominal} & {Net} & Governing static mode \\
			& {(N)} & {(mm)} & {(MPa)} & {(MPa)} & \\
			\midrule
			CNT00 & 35681 & 0.514 & 1784.0 & 2230.0 & matrix split with delayed local fibre activation (reported load level, overload continuation) \\
			CNT45 & 989 & 0.152 & 49.4 & 61.8 & off-axis matrix/inter-fibre crack \\
			CNT90 & 729 & 0.164 & 36.4 & 45.5 & transverse matrix/inter-fibre crack \\
			\bottomrule
		\end{tabularx}
	\end{table}
	
	\begin{figure}[tbp]
		\centering
		\includegraphics[height=0.82\textheight,keepaspectratio]{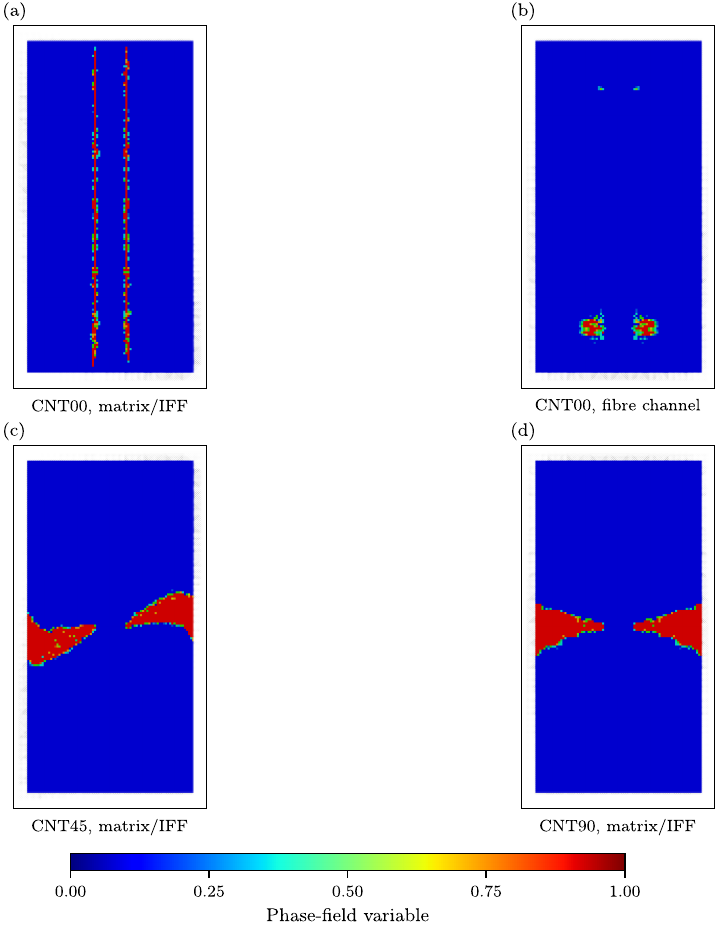}
		\caption{Static CNT crack-path benchmark represented by the mode-resolved phase-field variables. Panel (a) shows the matrix/inter-fibre phase-field \(\phi_{\mathrm{if}}\) for the \(0^\circ\) specimen, where longitudinal splitting develops from the notch. Panel (b) shows the corresponding delayed fibre-dominated phase-field activation, represented by \(\phi_{\mathrm{f}}\), under the overloaded \(0^\circ\) static continuation. Panels (c) and (d) show the governing matrix/inter-fibre phase-field for the \(45^\circ\) and \(90^\circ\) specimens, respectively, with off-axis cracking in the \(45^\circ\) case and transverse cracking in the \(90^\circ\) case. The contours are plotted with a common scale \(0\leq\phi\leq1\).}
		\label{fig:cnt-static-benchmark}
	\end{figure}
	
	\subsection{CNT fatigue response}
	
	The CNT fatigue results, summarized in Table~\ref{tab:cnt-fatigue}, show a strong orientation dependence while preserving the same material and fatigue card. CNT90 fails first by transverse matrix/inter-fibre cracking, CNT45 fails next by off-axis matrix/inter-fibre cracking, and CNT00 remains a runout case to $2.0\times10^{5}$ cycles. In CNT00, the matrix split forms and stabilizes, while the fibre phase field remains inactive throughout the fatigue analysis.
	
	\begin{table}[!htbp]
		\centering
		\small
		\caption{CNT fatigue results at $R=0.1$. Failure cycles are defined from the major load drop or collapse for failed cases. Runout is reported when no collapse occurs within the simulated cycle window.}
		\label{tab:cnt-fatigue}
		\begin{tabularx}{\textwidth}{p{0.12\textwidth}p{0.16\textwidth}p{0.22\textwidth}p{0.16\textwidth}X}
			\toprule
			Case & $U_{\max}$ (mm) & Representative life (cycles) & Final $RF/RF_0$ & Governing fatigue mode \\
			\midrule
			CNT00 & 0.073 & ${>}\,2.0\times10^{5}$ & 0.898 & stable matrix/inter-fibre split; no fibre crack \\
			CNT45 & 0.103 & $2.22\times10^{3}$ & collapse & off-axis matrix/inter-fibre crack \\
			CNT90 & 0.116 & $1.01\times10^{3}$ & collapse & transverse matrix/inter-fibre crack \\
			\bottomrule
		\end{tabularx}
	\end{table}
	
	\begin{figure}[tbp]
		\centering
		\includegraphics[height=0.80\textheight,keepaspectratio]{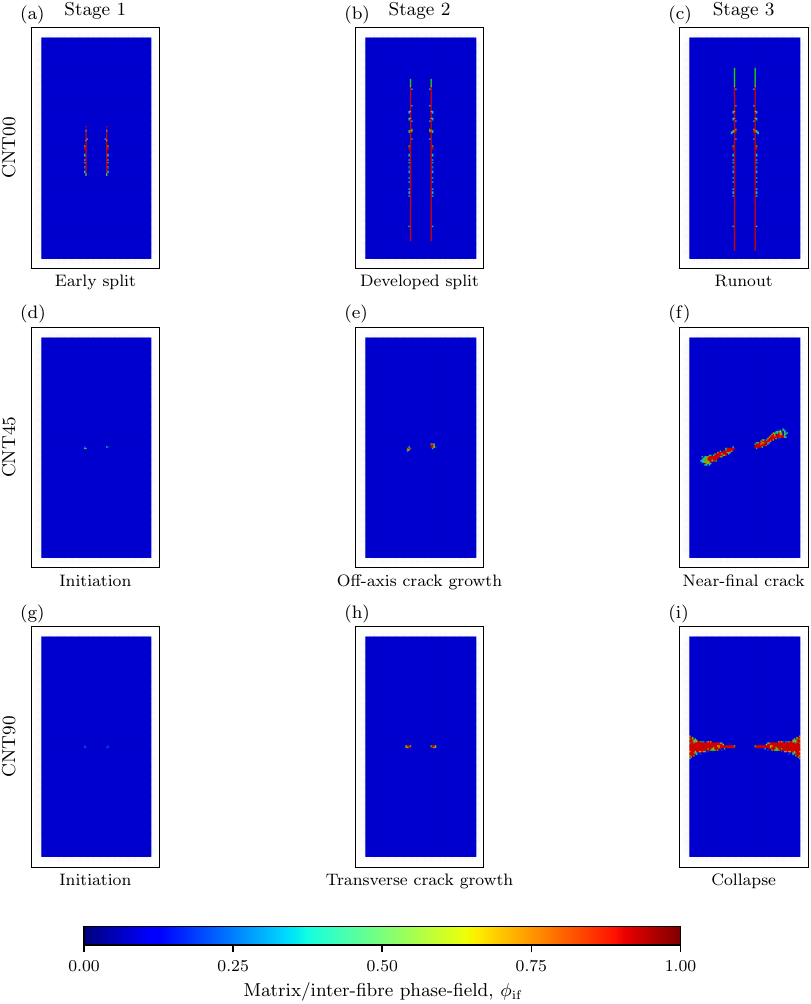}
		\caption{CNT fatigue evolution represented by the matrix/inter-fibre phase-field \(\phi_{\mathrm{if}}\). Each row corresponds to one fibre orientation and each column shows a representative stage of the fatigue process, selected to visualize initiation, crack growth, and the final crack-path state. The \(0^\circ\) specimen develops stable longitudinal matrix/inter-fibre splitting and remains a runout case at \(N=200{,}000\) cycles without fibre-dominated crack formation. The \(45^\circ\) specimen develops an off-axis matrix/inter-fibre crack, while the \(90^\circ\) specimen develops a transverse matrix/inter-fibre crack followed by rapid collapse. The contours are plotted with a common scale \(0\leq\phi_{\mathrm{if}}\leq1\), allowing direct comparison of the active damage mode and crack-path evolution across orientations.}
		\label{fig:cnt-fatigue-evolution}
	\end{figure}
	
	\section{Open-hole tension benchmark: a second geometry}\label{Open-hole}
	
	\subsection{Static OHT response}
	
	The OHT benchmark is a second, independent notched geometry. It tests whether the
	mode-resolved behaviour established on the CNT specimen is a robust property of the
	fixed card, or merely an artefact of the centred-notch configuration. The same
	material and fatigue card is used, with no geometry-specific tuning. The final OHT mesh uses $h=0.20$ mm and no O-grid refinement patch around the hole. The static results are summarized in Table~\ref{tab:oht-static}. OHT45 and OHT90 show clean matrix/inter-fibre dominated failures, with off-axis and transverse crack paths, respectively. OHT00 shows longitudinal matrix/inter-fibre splitting around the hole with localized fibre activation near the overload state. Raw OHT00 force spikes are rejected; the reported load level is the representative smoothed load level selected before the post-localization spike regime.
	
	\begin{table}[!htbp]
		\centering
		\small
		\caption{OHT static results on the final $h=0.20$ mm uniform Cartesian mesh. The $45^\circ$ and $90^\circ$ values are reported as strengths (same matrix/inter-fibre mode as CNT, consistent definition). The OHT00 value is a representative smoothed load level after rejecting nonphysical raw force spikes; it is not defined equivalently to the CNT $0^\circ$ value, and the two $0^\circ$ numbers document mode sequence and load scale rather than a quantitative slit-versus-hole notch-strength ranking.}
		\label{tab:oht-static}
		\begin{tabularx}{\textwidth}{l S[table-format=5.0] S[table-format=3.1] S[table-format=4.1] X}
			\toprule
			Case & {Reported force} & {Nominal} & {Net-section} & Governing static mode \\
			& {(N)} & {(MPa)} & {(MPa)} & \\
			\midrule
			OHT00 & 16482 & 824.1 & 1030.1 & matrix-first mixed longitudinal splitting (reported load level) \\
			OHT45 & 1066 & 53.3 & 66.6 & off-axis matrix/inter-fibre crack \\
			OHT90 & 777 & 38.9 & 48.6 & transverse matrix/inter-fibre crack \\
			\bottomrule
		\end{tabularx}
	\end{table}
	
	\begin{figure}[tbp]
		\centering
		\includegraphics[height=0.82\textheight,keepaspectratio]{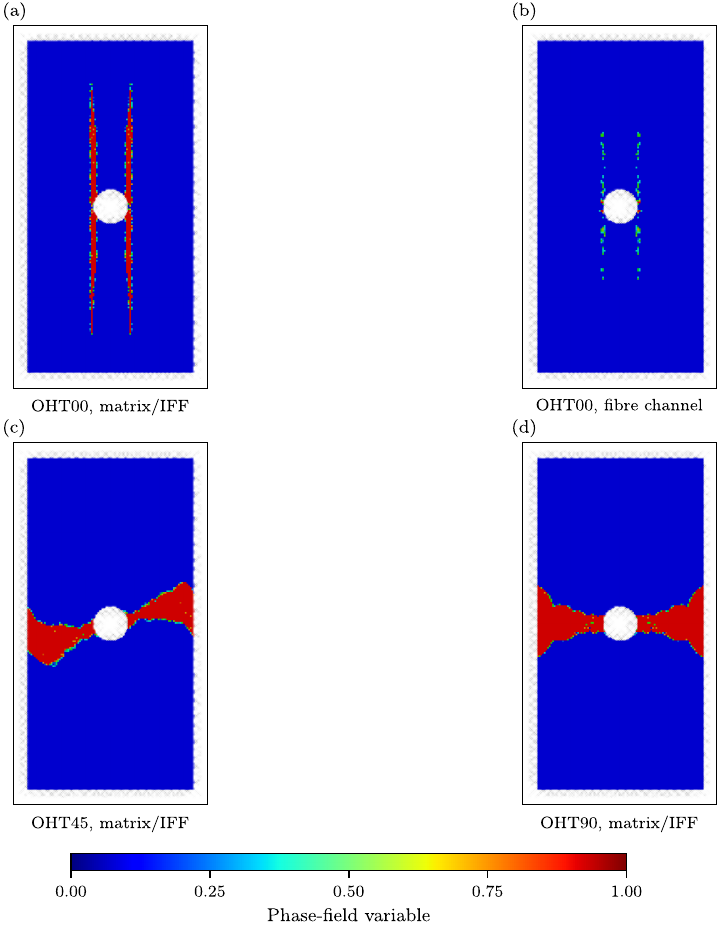}
		\caption{Static OHT crack-path benchmark represented by the mode-resolved phase-field variables. Panel (a) shows the matrix/inter-fibre phase-field \(\phi_{\mathrm{if}}\) for the \(0^\circ\) specimen, where longitudinal splitting develops around the hole. Panel (b) shows the corresponding fibre phase-field \(\phi_{\mathrm{f}}\), indicating localized fibre-dominated phase-field activation in the mixed \(0^\circ\) static failure process. Panels (c) and (d) show the governing matrix/inter-fibre phase-field for the \(45^\circ\) and \(90^\circ\) specimens, respectively, with off-axis cracking in the \(45^\circ\) case and transverse cracking in the \(90^\circ\) case. The contours are plotted with a common scale \(0\leq\phi\leq1\).}
		\label{fig:oht-static-benchmark}
	\end{figure}
	
	The computed transverse and off-axis static response levels are used here as numerical
	benchmark values, not as a calibrated experimental validation. Their order of magnitude
	is nonetheless consistent with published IM7/8552 lamina data. The NCAMP qualification
	report for Hexcel 8552/IM7 unidirectional prepreg reports a room-temperature-dry
	transverse tensile strength $F_{2tu}$ of $9.29$~ksi ($\approx 64$~MPa), together with
	in-plane shear strengths of $7.76$~ksi ($\approx 54$~MPa, 0.2\% offset) and $13.22$~ksi
	($\approx 91$~MPa, 5\% strain) \citep{marlett2011ncamp}. These values are consistent with
	the strength scale of the present $45^\circ$ and $90^\circ$ CNT and OHT matrix/inter-fibre
	responses, while remaining far below the longitudinal tensile strength scale. The
	comparison is intended only as a plausibility check on the active failure modes and the
	strength hierarchy; it is not a claim of quantitative experimental calibration. This
	cautious interpretation is consistent with later IM7/8552 transverse-strength studies,
	which show that transverse tensile strength depends on test geometry, specimen size, and
	surface condition \citep{arndt2020transverse}. Matrix-dominated transverse and shear
	behaviour, and the relevance of Puck-type matrix failure descriptions for IM7/8552 under
	combined transverse and shear loading, have also been reported by Koerber et al.
	\citep{koerber2010highstrain}.
	
	\subsection{OHT fatigue response}
	
	The OHT fatigue amplitudes are selected as 70\% of the static displacement at which the matrix/inter-fibre channel first reached $\phi_{\mathrm{if}}>0.10$ in the corresponding OHT static simulation. The resulting fatigue responses are summarized in Table~\ref{tab:oht-fatigue}. OHT90 fails at approximately $1.16\times10^{3}$ cycles by a transverse matrix/inter-fibre fatigue crack. OHT45 fails at approximately $1.57\times10^{3}$ cycles by an off-axis matrix/inter-fibre fatigue crack. OHT00 runs out to $2.0\times10^{5}$ cycles with a stable longitudinal matrix split, limited reaction-force degradation, and no fibre phase-field activation.
	
	\begin{table}[!htbp]
		\centering
		\small
		\caption{OHT displacement-controlled fatigue results at $R=0.1$.}
		\label{tab:oht-fatigue}
		\begin{tabularx}{\textwidth}{p{0.12\textwidth}p{0.17\textwidth}p{0.20\textwidth}p{0.20\textwidth}X}
			\toprule
			Case & $U_{\max}$ (mm) & Result (cycles) & Final/critical response & Governing fatigue mode \\
			\midrule
			OHT90 & 0.123725 & $N_f\approx1.16\times10^{3}$ & collapse & transverse matrix/inter-fibre crack \\
			OHT45 & 0.115891 & $N_f\approx1.57\times10^{3}$ & collapse & off-axis matrix/inter-fibre crack \\
			OHT00 & 0.095944 & $N_f>2.0\times10^{5}$ & final $RF/RF_0=0.898$ & stable longitudinal matrix/inter-fibre split; no fibre crack \\
			\bottomrule
		\end{tabularx}
	\end{table}
	
	\begin{figure}[tbp]
		\centering
		\includegraphics[height=0.80\textheight,keepaspectratio]{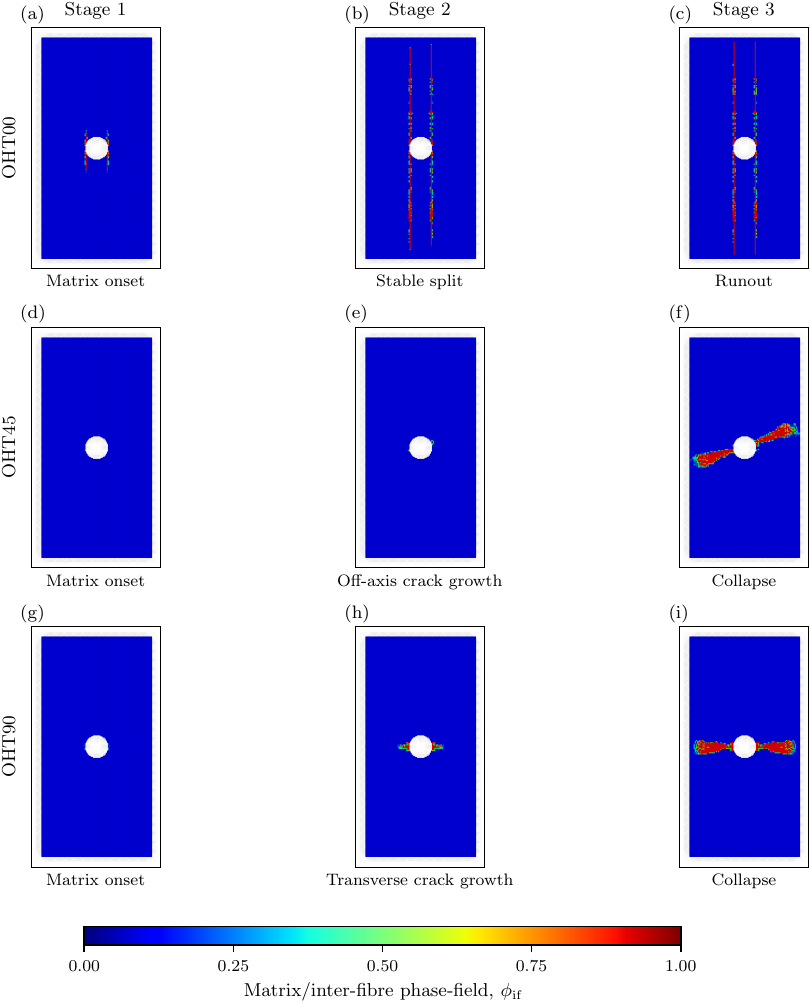}
		\caption{OHT fatigue evolution represented by the matrix/inter-fibre phase-field \(\phi_{\mathrm{if}}\). Each row corresponds to one fibre orientation and each column shows a representative stage of the fatigue process, selected to visualize matrix/inter-fibre onset, crack growth, and collapse or runout. The \(0^\circ\) specimen develops stable longitudinal matrix/inter-fibre splitting around the hole and remains a runout case at \(N=200{,}000\) cycles without fibre-dominated crack formation. The \(45^\circ\) specimen develops an off-axis matrix/inter-fibre crack followed by collapse, whereas the \(90^\circ\) specimen develops a transverse matrix/inter-fibre crack followed by abrupt loss of load-carrying capacity. The contours are plotted with a common scale \(0\leq\phi_{\mathrm{if}}\leq1\), allowing direct comparison of the active damage mode and crack-path evolution across orientations.}
		\label{fig:oht-fatigue-evolution}
	\end{figure}
	
	\subsection{Load-controlled OHT90 amplitude check}
	
	Displacement control is useful for stable crack-path verification, but lower-amplitude displacement-controlled fatigue can partially unload itself after stiffness loss. Therefore, a compact load-controlled OHT90 amplitude check is included as supplementary evidence. The maximum loads are 70\%, 60\%, and 50\% of the reference OHT90 static capacity. Under load control, failure is measured by compliance growth, $C/C_0$, where $C=U_{\mathrm{top}}/F_{\max}$ and $C_0$ is the initial compliance. A practical failure indicator is $C/C_0\ge2$. The results in Table~\ref{tab:oht-loadcontrol} show the expected trend: reducing load delays matrix initiation, compliance growth, and final failure while preserving the same transverse matrix/inter-fibre crack mode.
	
	\begin{table}[!htbp]
		\centering
		\small
		\caption{Additional OHT90 load-controlled amplitude check. Failure is defined by history-based compliance growth $C/C_0\approx2$. The governing mode is transverse matrix/inter-fibre cracking for all three load levels.}
		\label{tab:oht-loadcontrol}
		\begin{tabular}{l S[table-format=1.2] S[table-format=3.1] S[table-format=1.2e-1] S[table-format=1.2e-1]}
			\toprule
			Case & {$F_{\max}/F_{\mathrm{static}}$} & {$F_{\max}$ (N)} & {Matrix initiation} & {$N_f$ by $C/C_0\approx2$} \\
			& & & {(cycles)} & {(cycles)} \\
			\midrule
			LC70 & 0.70 & 544.0 & 1.14e3 & 1.21e3 \\
			LC60 & 0.60 & 466.3 & 1.70e3 & 2.03e3 \\
			LC50 & 0.50 & 388.6 & 3.31e3 & 1.02e4 \\
			\bottomrule
		\end{tabular}
	\end{table}
	
	\begin{figure}[tbp]
		\centering
		\includegraphics[width=\linewidth]{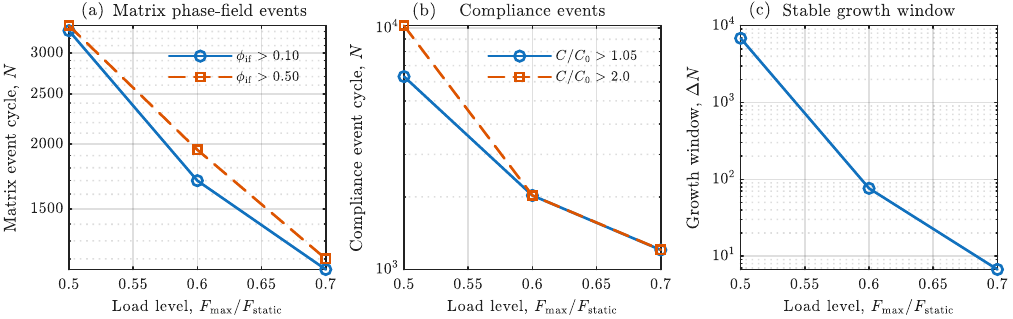}
		\caption{Load-control amplitude sensitivity for the OHT90 configuration. Panel (a) shows the cycles at which the matrix/inter-fibre phase-field reaches \(\phi_{\mathrm{if}}>0.10\) and \(\phi_{\mathrm{if}}>0.50\). Panel (b) shows compliance-growth events based on \(C/C_0>1.05\) and \(C/C_0>2.0\). Panel (c) shows the growth window \(\Delta N=N_{C/C_0>2.0}-N_{\phi_{\mathrm{if}}>0.50}\), highlighting that the lower load level permits a longer stable matrix/inter-fibre damage-development stage before rapid compliance growth.}
		\label{fig:d6b-loadcontrol}
	\end{figure}
	
	Representative phase-field contour fields for the OHT90 load-control amplitude study are provided in Figure~\ref{fig:supp-d6b-oht90-loadcontrol-contours} of the Supplementary Information.
	
	\subsection{Geometric OHT90 hole-size sensitivity}
	\label{sec:oht-hole-size-sensitivity}
	
	A final compact D7 check tests whether the same OHT90 model reacts consistently to geometric notch severity. The hole diameter is varied while keeping the specimen width $W=20$ mm, orientation, material card, fatigue card, mesh philosophy, and load-control definition unchanged. The selected diameters are $D=2.5$, 4.0, and 5.5 mm, corresponding to $D/W=0.125$, 0.200, and 0.275. The $D=4.0$ mm case is the reference OHT90 geometry. The new $D=2.5$ and $D=5.5$ mm meshes use the same $h=0.20$ mm projected-hole topology, giving approximately 12.5 and 27.5 elements across the hole diameter, respectively.
	
	The static trend is monotonic and physically consistent. Increasing the hole diameter decreases the smoothed peak force from 871.6 N to 661.8 N and decreases the nominal strength from 43.6 MPa to 33.1 MPa. The net-section strength also decreases moderately, from 49.8 MPa to 45.6 MPa, reflecting increased notch severity rather than only reduced ligament area. In all three cases, the governing static mode remains transverse matrix/inter-fibre cracking; fibre activation appears only after peak collapse and is not the controlling event.
	
	For the fatigue part, all three hole sizes are subjected to the same gross load-controlled cycle, $F_{\max}=466.327$ N with $R=0.1$. This is the LC60 load level of the reference $D=4.0$ mm OHT90 case. Under this identical applied load, the matrix/inter-fibre fatigue life decreases strongly with increasing hole diameter. The small-hole case reaches matrix phase-field onset at approximately $2.81\times10^{3}$ cycles and compliance-based failure at $2.87\times10^{3}$ cycles. The reference $D=4.0$ mm case fails at approximately $2.03\times10^{3}$ cycles. The large-hole case reaches matrix cracking and rapid compliance growth at approximately $9.14\times10^{2}$ cycles. Table~\ref{tab:oht-geometry} summarizes the static and fatigue trends.
	
	\begin{table}[!htbp]
		\centering
		\small
		\caption{D7 OHT90 geometric notch-size sensitivity. Static response is evaluated from the smoothed peak load. Fatigue response uses the same gross load-controlled cycle for all hole sizes, $F_{\max}=466.327$ N and $R=0.1$. Failure is defined by history-based compliance growth $C/C_0\approx2$.}
		\label{tab:oht-geometry}
		\begin{tabular}{S[table-format=1.1] S[table-format=1.3] S[table-format=3.1] S[table-format=2.1] S[table-format=2.1] S[table-format=1.2e-1] l}
			\toprule
			{$D$ (mm)} & {$D/W$} & {Peak (N)} & {Nom. (MPa)} & {Net (MPa)} & {Fatigue $N_f$} & Mode \\
			\midrule
			2.5 & 0.125 & 871.6 & 43.6 & 49.8 & 2.87e3 & transverse inter-fibre \\
			4.0 & 0.200 & 777.2 & 38.9 & 48.6 & 2.03e3 & transverse inter-fibre \\
			5.5 & 0.275 & 661.8 & 33.1 & 45.6 & 9.14e2 & transverse inter-fibre \\
			\bottomrule
		\end{tabular}
	\end{table}
	
	Representative static and fatigue phase-field contour fields for the OHT90 hole-size sensitivity study are provided in Figure~\ref{fig:supp-oht90-geometry-contours} of the Supplementary Information.
	
	\begin{figure}[tbp]
		\centering
		\includegraphics[width=\linewidth]{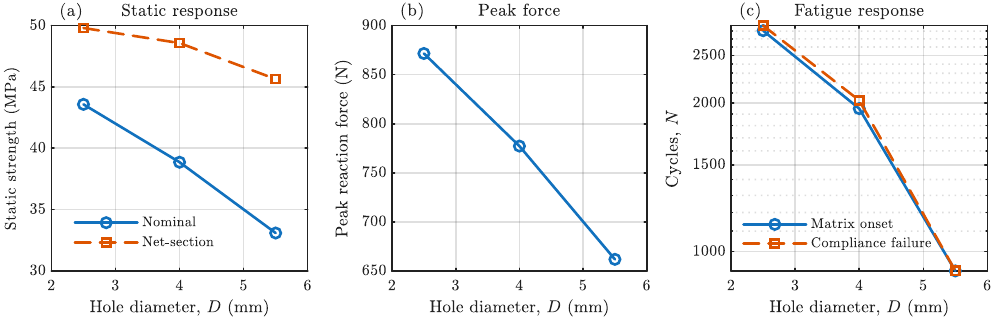}
		\caption{Open-hole geometry sensitivity for the OHT90 configuration. Panel (a) compares the nominal and net-section static strengths for hole diameters \(D=2.5\), \(4.0\), and \(5.5\) mm. Panel (b) shows the corresponding smoothed static peak reaction force. Panel (c) compares the matrix phase-field onset cycle and the compliance-based fatigue failure cycle under the same load-controlled fatigue level, \(F_{\max}=466.327\) N and \(R=0.1\). Increasing the hole diameter reduces the static load capacity and accelerates matrix/inter-fibre fatigue localization and collapse, while the governing crack mode remains transverse matrix/inter-fibre cracking.}
		\label{fig:d7-geometry-sensitivity}
	\end{figure}
	
		\section{Discussion}\label{Discussion}
	\begin{revblock}
	\subsection{Split-induced shielding as a mechanism, not a numerical artefact}
	The most important structural result is the fibre-aligned response. In both the CNT and OHT \(0^\circ\) configurations, the notch does not immediately trigger a fibre-dominated crack. It first activates the matrix/inter-fibre channel, producing longitudinal splits that run approximately parallel to the fibres. This sequence is physically significant. A longitudinal split damages the matrix/inter-fibre load-transfer path, but it does not cut the primary fibre load path. Instead, it relaxes the transverse and shear tractions that concentrate around the notch or hole. In local ply variables, the split reduces the stress components that enter the matrix/inter-fibre driving energy, \(\sigma_2\) and \(\sigma_{12}\), while the continuous fibres can still carry the main longitudinal stress \(\sigma_1\) over the remaining ligament.
	
	This interpretation is supported by direct field extraction from the accepted \(0^\circ\) fatigue analyses. Robust \(99\)th-percentile values were evaluated over a central near-notch region rather than at single integration points, so that the reported quantities are not controlled by isolated element extrema. In CNT00, the matrix/inter-fibre phase field reaches \(\max\phi_{\mathrm{if}}=1\) after split formation and remains saturated at runout, while the fibre phase field remains \(\max\phi_{\mathrm{f}}=0\) throughout the fatigue analysis. Over the same history, the transverse and shear stress measures associated with the matrix/inter-fibre channel decrease strongly: the transverse indicator \(P_{99}(\sigma_{\perp})\) falls from \(\SI{12.25}{\mega\pascal}\) at the start of fatigue to \(\SI{2.89}{\mega\pascal}\) at \(N=2\times10^{5}\), a reduction of about \(76\%\), and the in-plane shear indicator \(P_{99}(\tau_{12})\) falls from \(\SI{40.85}{\mega\pascal}\) to \(\SI{7.51}{\mega\pascal}\), a reduction of about \(82\%\). The fibre-direction stress remains substantial, \(P_{99}(\sigma_{\parallel})=\SI{349.31}{\mega\pascal}\) at runout, confirming that the fibres continue to carry load. Nevertheless, the fibre effective effort stays far below activation, decreasing from \(P_{99}(F_{\mathrm{f}}^{\mathrm{eff}})=0.206\) at fatigue start to \(0.151\) at runout. The open-hole case shows the same behaviour more strongly: \(P_{99}(\sigma_{\perp})\) decreases from \(\SI{31.32}{\mega\pascal}\) to \(\SI{2.64}{\mega\pascal}\) and \(P_{99}(\tau_{12})\) from \(\SI{51.05}{\mega\pascal}\) to \(\SI{2.29}{\mega\pascal}\), while \(P_{99}(F_{\mathrm{f}}^{\mathrm{eff}})\) decreases from \(0.286\) to \(0.181\) and the fibre-direction stress remains finite at \(P_{99}(\sigma_{\parallel})=\SI{421.68}{\mega\pascal}\). At runout, the fibre equivalent fatigue driver is also below the fibre threshold \(F_{\mathrm{th},\mathrm{f}}=0.12\), with \(P_{99}(\widehat{F}_{\mathrm{f}})=0.068\) for CNT00 and \(0.082\) for OHT00. The fibre channel is therefore subcritical both mechanically, with effective effort far below unity, and in fatigue at runout, with the equivalent driver below threshold. The field-extracted metrics are collected in Supplementary Table~\ref{tab:shielding-metrics}.
	
	The model therefore predicts a two-step redistribution. First, the low-resistance matrix/inter-fibre channel localizes, because its strength, fracture energy, and fatigue threshold are much smaller than those of the fibre channel. Second, once a stable longitudinal split exists, the local notch constraint is reduced and the fibre Puck effort remains below activation. The fibre channel is consequently \emph{shielded}: not because the fibres are unloaded, but because the local multiaxial concentration that would drive fibre-channel activation is blunted by matrix/inter-fibre splitting. \rev{The extraction quantifies exactly this distinction: the fibre-direction stress remains of the order of several hundred megapascals at runout, whereas the transverse and shear measures are reduced by roughly \(76\)--\(96\%\). The split therefore removes much of the transverse/shear notch constraint without removing the longitudinal fibre load path.} This distinction is essential. The \(0^\circ\) runout is not an undamaged state; it is a stable damaged state in which \(\phi_{\mathrm{if}}\) has formed a longitudinal split while \(\phi_{\mathrm{f}}\) remains inactive.
	
	This interpretation is consistent with experimental observations of longitudinal splitting in notched unidirectional composites. Centre-notched unidirectional graphite/epoxy tests under tensile loading have shown longitudinal splitting as a characteristic damage mode \citep{wolla1987splitting}. Open-hole unidirectional composite tests have likewise reported fibre-direction splits growing from the hole under longitudinal tension \citep{bazhenov1998splitting}. The stress-redistribution aspect is also consistent with analyses showing that fibre-direction splits at notch tips can blunt the notch and substantially reduce the local stress concentration under remote tension \citep{liu2016stress}. The present formulation does not claim that experiments directly measure the phase-field variables or the fibre-channel Puck effort. Rather, it provides a model-level interpretation of an experimentally observed splitting/blunting mechanism: the matrix/inter-fibre phase-field band represents the split, and the suppressed fibre phase field represents the corresponding shielding of the fibre-dominated fatigue channel.
	
	The static overload continuations are important for the same reason. They show that the fibre channel is not artificially disabled. Under the selected fatigue amplitudes, the split is sufficient to keep the fibre effort subcritical through \(2\times10^{5}\) cycles. Under sustained monotonic overload, however, the average longitudinal stress continues to rise after the split has formed. Once the split can no longer relieve the increasing fibre-direction demand, localized fibre-channel activation appears. Thus the fatigue runout and the static delayed-fibre response are not contradictory; they are two parts of the same hierarchy: matrix/inter-fibre splitting first, fibre activation only after the shielding capacity is exhausted.
	
	\subsection{Consistency across orientation, geometry, load level, and notch severity}
	The CNT and OHT simulations should be read as independent mechanism tests rather than as a calibration sequence. The same material and fatigue card is used for all orientations and both geometries. No parameter is changed to make a specific coupon fail by a desired mode or at a desired cycle count. The significance of the structural results is therefore not the absolute agreement of any one life value with an experiment, but the repeated emergence of the same physical hierarchy under distinct boundary-value problems. Transverse $90^\circ$ loading produces transverse matrix/inter-fibre cracking; off-axis $45^\circ$ loading produces inclined matrix/inter-fibre cracking; fibre-aligned $0^\circ$ loading produces longitudinal splitting and delayed or absent fibre activation.
	
	The supplementary load-controlled and hole-size studies provide two additional checks on this hierarchy. Reducing the OHT90 load level delays both matrix phase-field onset and compliance-based failure, while the crack mode remains transverse matrix/inter-fibre cracking. Increasing the hole diameter under the same gross load-controlled fatigue cycle accelerates the same transverse matrix/inter-fibre localization and failure. These trends are important because they perturb the driving force in two different ways: one changes the applied amplitude, the other changes the local stress concentration. In both cases the model responds through the expected local phase-field evolution rather than through any change in the material card. This supports the interpretation that the formulation captures a mechanism-level response, not a geometry-specific numerical pattern.
	
	The comparison between CNT and OHT must nevertheless be kept precise. The $45^\circ$ and $90^\circ$ static values are governed by the same matrix/inter-fibre failure mode in both geometries and are extracted under consistent definitions, so their strength scale can be meaningfully compared. The $0^\circ$ static values serve a different purpose. They document the mode sequence and the load level at which delayed fibre activation appears after longitudinal splitting; they are not equivalent ultimate-strength definitions for a slit and a circular hole at fixed net ligament. This distinction prevents the $0^\circ$ overload continuations from being misread as a notch-strength ranking.
	
	\subsection{Physical meaning and limits of the phase-field bands}\label{sec:bands}
	The plotted phase-field cracks are regularized fracture/process zones. Their apparent width is controlled by the length scale and by mesh resolution, not by a directly measurable crack opening. This point matters particularly for matrix-dominated composite failure, where the physical damage zone may contain microcracks, fibre--matrix debonding, shear deformation, and local coalescence rather than a single sharp mathematical discontinuity. The useful information in the contours is therefore the active channel, crack path, onset sequence, and relation to structural force or compliance response.
	
	The reported contours satisfy the numerical requirements for this interpretation. The CNT mesh gives $\ell_{\mathrm{if}}/h=4$ and the OHT mesh gives $\ell_{\mathrm{if}}/h=5$, so the matrix/inter-fibre regularization band is resolved by several elements. The OHT mesh comparison preserves the governing crack mechanisms, and the cycle-block check preserves both the fatigue lives and the crack paths within the reported tolerance. Consequently, the band-like contours should be interpreted as length-scale-controlled representations of the fracture process zone, not as unresolved mesh artefacts.
	
	\subsection{Numerical observables and structural failure definitions}\label{sec:spikes}
	Near localization, raw reaction-force or raw channel-effort spikes can appear because of solver cutbacks, abrupt local stiffness changes, and post-critical numerical instability. Such spikes are not material properties. The physically admissible observables are the phase-field pattern, the smooth part of the load--displacement or compliance response, and the consistency of the active channel with the surrounding stress state. For static displacement-controlled cases, reported values are therefore based on representative smoothed load levels and corresponding phase-field states. For displacement-controlled fatigue, structural failure is defined by major reaction-force loss together with phase-field crack formation. For load-controlled fatigue, force is prescribed and failure is more naturally identified by compliance growth.
	
	This convention is not a post-processing convenience; it is part of the physical interpretation of regularized fracture. A single-increment force spike without a corresponding stable phase-field pattern does not represent a measurable strength. Conversely, a stable phase-field split without global collapse, as in the $0^\circ$ fatigue cases, is a real damage state but not a structural failure. The same criterion is applied to all CNT and OHT cases, which is essential for comparing mechanisms across geometries.
	
	\subsection{Mean-stress scope and calibration boundary}
	The mean-stress term is included because polymer-composite fatigue is generally sensitive to tensile mean load: a higher tensile mean can keep matrix cracks more open, reduce closure-type shielding, and accelerate microcrack or debonding growth. The D2b one-element study confirms that the implemented term has the intended qualitative effect. When $\beta_{\mathrm{mean},\mathrm{if}}=0$, the equivalent fatigue driver remains essentially amplitude controlled; when $\beta_{\mathrm{mean},\mathrm{if}}=0.35$, increasing $R$ increases the equivalent inter-fibre driver and accelerates resistance degradation.
	
	The structural CNT and OHT benchmarks deliberately do not use this term for calibration. They are restricted to $R=0.1$ with $\beta_{\mathrm{mean},\mathrm{f}}=\beta_{\mathrm{mean},\mathrm{if}}=0$ so that the mode-resolved mechanism can be assessed without introducing an additional stress-ratio fit. A quantitative stress-ratio model for IM7/8552 would require experimental data at multiple $R$ ratios, and would be a separate identification problem rather than a consequence of the present numerical consistency study.
	
	\subsection{Limitations and path to experimental validation}
	The present work is a homogenized ply-level demonstration. It does not explicitly resolve individual fibres, fibre--matrix interfaces, matrix ligaments, pull-out, microvoids, or ply-to-ply interactions. These mechanisms are represented through effective channel-wise elastic, strength, fracture, and fatigue parameters. The implementation is elastic at ply scale; matrix plasticity, rate-dependent matrix dissipation, residual stresses, and variable-amplitude sequence effects remain outside the present scope. The current parameter card is used to verify numerical consistency, mechanism separation, and cross-geometry generality. It is not presented as an experimentally identified IM7/8552 fatigue-life model.
	
	These limitations define a clear validation path. The most direct experimental tests would combine notched UD CNT and OHT coupons with full-field strain measurement and post-mortem or in-situ damage mapping. The key observables would not be only final life, but the sequence of damage mechanisms: onset of longitudinal splitting in $0^\circ$ coupons, persistence or suppression of fibre fracture under fixed-amplitude cycling, off-axis crack growth in $45^\circ$ coupons, transverse matrix/inter-fibre cracking in $90^\circ$ coupons, and the effect of hole size or load level on compliance growth. Such data would allow the present fixed demonstration card to be replaced by an experimentally identified fatigue card and would test the split-induced shielding hypothesis quantitatively. Until that campaign is performed, the contribution of this paper is deliberately framed as a verified mode-resolved formulation and a cross-geometry consistency assessment, not as a calibrated life-prediction model.
	\end{revblock}

	\section{Conclusions}\label{Conclusions}
	
	A Puck-informed mode-resolved phase-field fatigue framework for UD composites has been developed, implemented, verified, and exercised on CNT and OHT benchmarks. The main conclusions are:
	
	\begin{enumerate}[leftmargin=0.7cm]
		\item Fatigue degradation in UD composites should not be represented by a single global scalar damage variable when fibre and matrix/inter-fibre mechanisms are both relevant. A two-channel description gives direct physical interpretability.
		\item The proposed framework separates fatigue accumulation from stiffness loss. Fatigue histories reduce the resistance of the corresponding channel, while the local phase fields $\phif$ and $\phiif$ govern actual stiffness degradation and crack-path evolution.
		\item One-element tests verify selective channel activation, and parameter sweeps confirm the expected roles of the accumulation coefficient, threshold, exponent, transition history, and degradation-shape parameter. This establishes controllability before structural benchmarks are interpreted.
		\item A fixed-amplitude one-element mean-stress sensitivity check demonstrates the optional channel-wise mean-stress capability. With $\beta_{\mathrm{mean},\mathrm{if}}=0$, the equivalent fatigue driver remains essentially independent of load ratio; with $\beta_{\mathrm{mean},\mathrm{if}}=0.35$, increasing tensile mean level increases the equivalent driver and accelerates inter-fibre degradation.
		\item CNT static and fatigue simulations at $0^\circ$, $45^\circ$, and $90^\circ$ reproduce distinct crack topologies using one fixed material and fatigue card: delayed fibre activation after longitudinal matrix splitting in CNT00, off-axis matrix/inter-fibre cracking in CNT45, and transverse matrix/inter-fibre cracking in CNT90.
		\item Applying the identical fixed card to a second, independent geometry (OHT) reproduces the same orientation-dependent mechanisms found in CNT, confirming that the predicted behaviour is a property of the formulation rather than of one notch. The $45^\circ$ and $90^\circ$ cases, which share a matrix/inter-fibre failure mode in both geometries, also agree in strength scale; the $0^\circ$ comparison is restricted to mode sequence and load scale rather than ultimate strength.
		\item The fibre-aligned runout is explained by split-induced shielding: longitudinal matrix splitting relieves the notch stress concentration and keeps the fibre channel below activation, so both $0^\circ$ configurations survive to $2.0\times10^{5}$ cycles while the matrix-dominated orientations fail within $\sim 10^{3}$ cycles.
		\item Supplementary load-controlled and geometric checks reproduce the expected trends: reducing $F_{\max}$ from 70\% to 50\% of static capacity delays matrix initiation and compliance-based failure, and increasing the hole diameter from 2.5 to 5.5 mm lowers static capacity and reduces compliance-based fatigue life from approximately $2.87\times10^{3}$ to $9.14\times10^{2}$ cycles, while the transverse matrix/inter-fibre mode is preserved throughout.
		\item Mesh/length-scale and cycle-block convergence checks support the robustness of the reported crack paths and lives. The regularized crack bands are length-scale controlled and should be interpreted as phase-field fracture/process zones rather than physical crack thicknesses.
	\end{enumerate}
	
	Overall, a single fixed mode-resolved fatigue card reproduces orientation-,
	load-level-, and notch-size-dependent fatigue mechanisms consistently across two
	independent notched geometries. This generality, obtained without per-configuration
	tuning, establishes the framework as a physically interpretable basis for future
	experimental calibration, laminate-level extension, and variable-amplitude loading.
	
	\clearpage
	\section*{Supplementary Information}
	
	% Store main-manuscript table counter before supplementary tables
	\newcounter{mainTableBeforeSI}
	\setcounter{mainTableBeforeSI}{\value{table}}
	
	% Supplementary table numbering
	\setcounter{table}{0}
	\renewcommand{\thetable}{S\arabic{table}}
	
	\begin{table}[H]
		\centering
		\color{black}
		\scriptsize
		\setlength{\tabcolsep}{4pt}
		\caption{Field-extracted shielding metrics for the fibre-aligned fatigue runout cases. Values are robust \(99\)th-percentile quantities over a central near-notch region, except for the phase-field maxima. The symbols \(\sigma_{\parallel}\), \(\sigma_{\perp}\), and \(\tau_{12}\) denote fibre-direction, transverse, and in-plane shear stress indicators, respectively, extracted from the degraded stress field in the \(0^\circ\) configurations.}
		\label{tab:shielding-metrics}
		\begin{tabular}{llccccccc}
			\toprule
			Case & Stage
			& \(\max\phi_{\mathrm{if}}\)
			& \(\max\phi_{\mathrm{f}}\)
			& \(P_{99}(\sigma_{\perp})\)
			& \(P_{99}(\tau_{12})\)
			& \(P_{99}(\sigma_{\parallel})\)
			& \(P_{99}(F_{\mathrm{f}}^{\mathrm{eff}})\)
			& \(P_{99}(\widehat{F}_{\mathrm{f}})\) \\
			& & & & MPa & MPa & MPa & -- & -- \\
			\midrule
			CNT00 & start  & 0.000 & 0.000 & 12.25 & 40.85 & 478.14 & 0.206 & 0.000 \\
			CNT00 & runout & 1.000 & 0.000 &  2.89 &  7.51 & 349.31 & 0.151 & 0.068 \\
			OHT00 & start  & 0.000 & 0.000 & 31.32 & 51.05 & 665.39 & 0.286 & 0.000 \\
			OHT00 & runout & 1.000 & 0.000 &  2.64 &  2.29 & 421.68 & 0.181 & 0.082 \\
			\bottomrule
		\end{tabular}
	\end{table}
	\FloatBarrier
	
	\setcounter{figure}{0}
	\renewcommand{\thefigure}{S\arabic{figure}}
	
	\subsection*{S1. Load-control contour fields for the OHT90 amplitude study}
	
\begin{figure}[H]
		\centering
		\includegraphics[height=0.66\textheight,keepaspectratio]{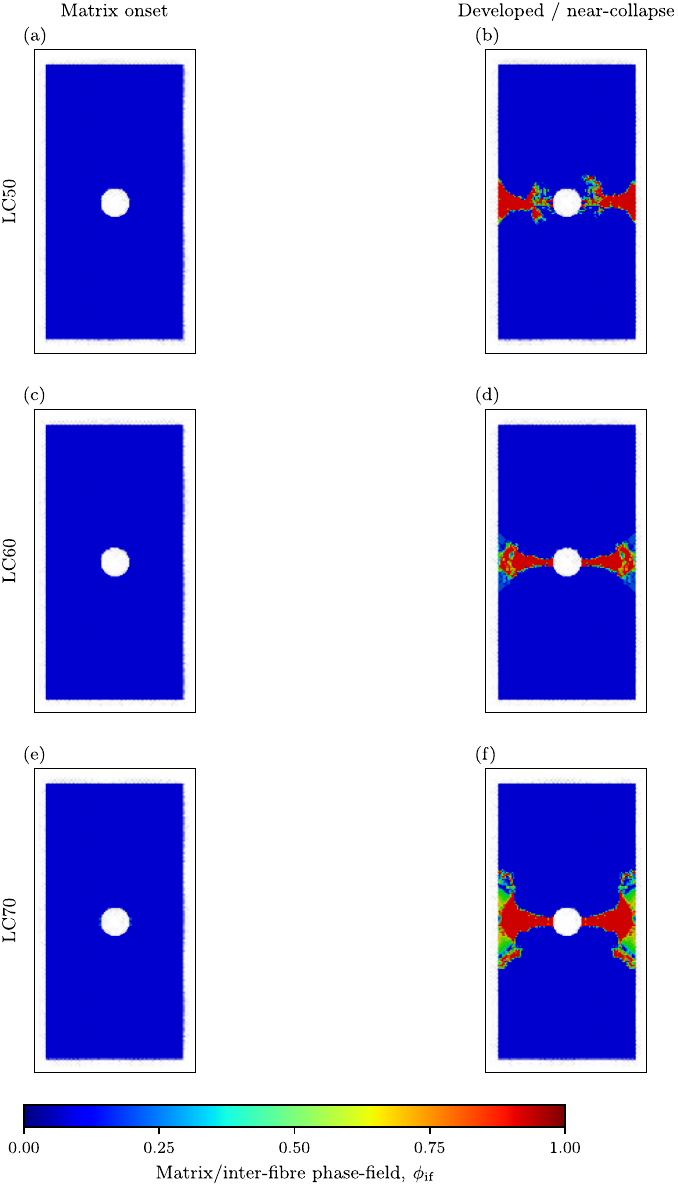}
		\caption{Supplementary load-control fatigue contour fields for the OHT90 configuration. The matrix/inter-fibre phase-field \(\phi_{\mathrm{if}}\) is shown for the LC50, LC60, and LC70 load levels. The left column shows the matrix/inter-fibre onset state, while the right column shows the developed or near-collapse state. All three load levels produce the same governing transverse matrix/inter-fibre crack mode, whereas increasing load level accelerates localization and collapse. The contours are plotted with a common scale \(0\leq\phi_{\mathrm{if}}\leq1\).}
		\label{fig:supp-d6b-oht90-loadcontrol-contours}
	\end{figure}

	Figure~\ref{fig:supp-d6b-oht90-loadcontrol-contours} provides representative phase-field contour fields for the OHT90 load-control amplitude study. The figure complements the quantitative load-control comparison in the main manuscript by showing that the governing fatigue damage mode remains transverse matrix/inter-fibre cracking for all investigated load levels.
	
	\subsection*{S2. Hole-size sensitivity contour fields for the OHT90 geometry study}
	
\begin{figure}[H]
		\centering
		\includegraphics[height=0.66\textheight,keepaspectratio]{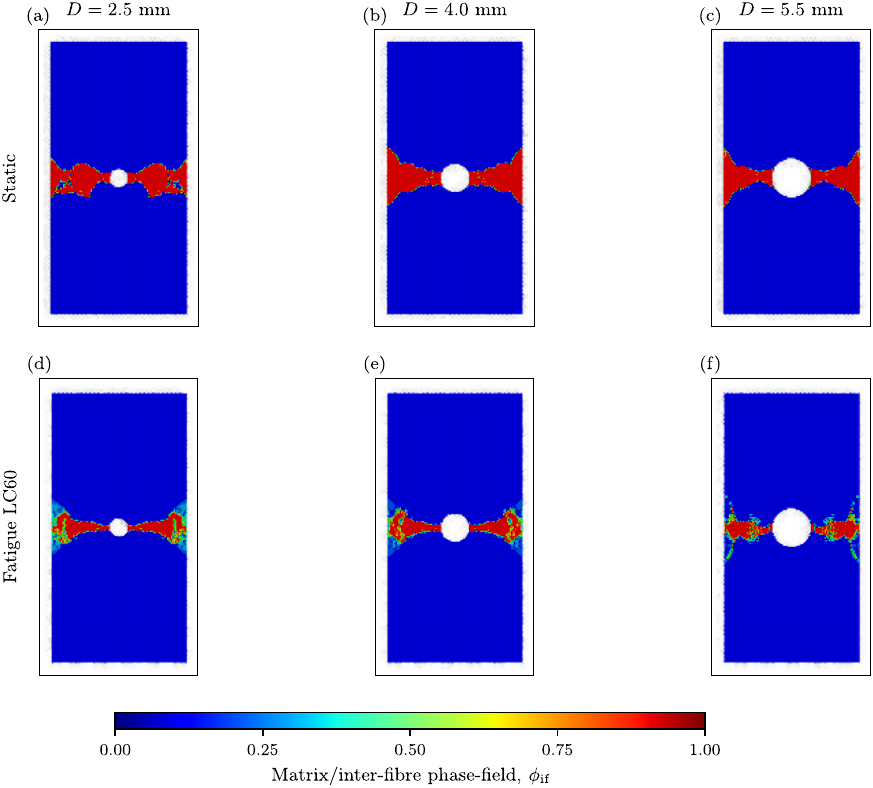}
		\caption{Supplementary hole-size sensitivity contour fields for the OHT90 configuration. The matrix/inter-fibre phase-field \(\phi_{\mathrm{if}}\) is shown for hole diameters \(D=2.5\), \(4.0\), and \(5.5\) mm. The top row shows representative static contours, while the bottom row shows the corresponding load-controlled fatigue contours at \(F_{\max}=466.327\) N and \(R=0.1\). Increasing the hole diameter reduces the ligament width and promotes stronger localization, while the governing failure mode remains transverse matrix/inter-fibre cracking. All contours are plotted with a common scale \(0\leq\phi_{\mathrm{if}}\leq1\).}
		\label{fig:supp-oht90-geometry-contours}
	\end{figure}

	Figure~\ref{fig:supp-oht90-geometry-contours} provides representative matrix/inter-fibre phase-field contours for the OHT90 hole-size sensitivity study. The figure complements the quantitative static-strength and fatigue-life trends reported in the main manuscript by showing that the governing crack mode remains transverse matrix/inter-fibre cracking for all investigated hole diameters.
	
	\FloatBarrier
	
	% Restore main table and figure numbering after supplementary material
	\setcounter{table}{\value{mainTableBeforeSI}}
	\renewcommand{\thetable}{\arabic{table}}
	\renewcommand{\thefigure}{\arabic{figure}}
	
	\clearpage
	\section*{Acknowledgements}
	The author gratefully acknowledges Dr. Pavan K. Asur Vijaya Kumar for valuable scientific discussions on Puck-informed multi-phase-field modelling and for his helpful insights.
	
	\section*{Data availability}
	The UMAT--UEL implementation files, representative Abaqus input decks, post-processing scripts, and processed result tables will be made available in a public repository or as supplementary material upon acceptance, subject to institutional approval.
	
	\section*{Declaration of competing interest}
	The author declares no known competing financial interests or personal relationships that could have appeared to influence the work reported in this paper.
	
	\section*{Declaration of generative AI and AI-assisted technologies in the manuscript preparation process}
	During the preparation of this work, the author used AI-assisted tools for language editing and grammar checking. After using these tools, the author reviewed and edited the content as needed and takes full responsibility for the content of the publication.
	
	\appendix
	\section{Nomenclature}
	
	\begin{longtable}{p{0.22\textwidth}p{0.70\textwidth}}
		\caption{Main symbols used in the manuscript.}\\
		\toprule
		Symbol & Description \\
		\midrule
		\endfirsthead
		\toprule
		Symbol & Description \\
		\midrule
		\endhead
		$\bm{u}$ & displacement vector \\
		$\bm{\varepsilon}$ & small-strain tensor \\
		$\bm{\sigma}$ & Cauchy stress tensor \\
		$\phif$ & fibre-dominated phase-field variable \\
		$\phiif$ & matrix/inter-fibre phase-field variable \\
		$G_{\mathrm{c,f}}$, $G_{\mathrm{c,if}}$ & fibre and inter-fibre fracture energies \\
		$\ell_{\mathrm{f}}$, $\ell_{\mathrm{if}}$ & fibre and inter-fibre phase-field length scales \\
		$\bm{A}_{\mathrm{f}}$, $\bm{A}_{\mathrm{if}}$ & channel-wise anisotropic structural tensors (crack-orientation projectors) \\
		$c_w$ & crack-surface normalization constant ($c_w=2$ for AT2) \\
		$\kappaf$, $\kappaif$ & accumulated fibre and inter-fibre fatigue histories \\
		$\chif$, $\chiif$ & fibre and inter-fibre fatigue resistance degradation functions \\
		$C_{\mathrm{fat},i}$ & channel-wise fatigue accumulation coefficient \\
		$p_{\mathrm{fat},i}$ & channel-wise fatigue exponent \\
		$F_{\mathrm{th},i}$ & channel-wise fatigue threshold \\
		$\kappa_{T,i}$ & channel-wise transition fatigue history for resistance degradation \\
		$a_i$ & channel-wise post-threshold degradation shape parameter \\
		$F^{\mathrm{raw}}_i$ & raw Puck-informed channel effort \\
		$F^{\mathrm{eff}}_i$ & fatigue-modified effective channel effort \\
		$F_{a,i}$, $F_{m,i}$ & channel-wise cycle amplitude and mean efforts \\
		$\widehat{F}_i$ & equivalent (mean-corrected) fatigue driver \\
		$\widehat{F}_{\max}$ & upper cap on the equivalent fatigue driver (numerical robustness) \\
		$\epsilon$ & lower bound on the mean-stress denominator \\
		$U_a$ & cyclic displacement amplitude used in the mean-stress sensitivity study \\
		$\beta_{\mathrm{mean},i}$ & channel-wise mean-stress sensitivity parameter \\
		$D$ & open-hole diameter in the OHT geometric sensitivity study \\
		$W$ & specimen width \\
		$F_{\max}$ & maximum applied force in load-controlled fatigue \\
		$R$ & cyclic load or displacement ratio \\
		$\Delta N$ & accepted block of fatigue cycles \\
		$C/C_0$ & compliance ratio used for load-controlled fatigue failure \\
		\bottomrule
	\end{longtable}

\end{document}